\documentclass[12pt]{article}

\usepackage{amsmath}
\usepackage{amssymb}
\usepackage{url}
\usepackage{hyperref}
\usepackage{xspace}
\usepackage{graphicx}

\usepackage{a4wide}

\newcommand{\be}{\begin{equation}}
\newcommand{\ee}{\end{equation}}
\newcommand{\bea}{\begin{eqnarray}}
\newcommand{\eea}{\end{eqnarray}}
\newcommand{\bi}{\begin{itemize}}
\newcommand{\ei}{\end{itemize}}
\newcommand{\ben}{\begin{enumerate}}
\newcommand{\een}{\end{enumerate}}

\newcommand{\lp}{\left(}
\newcommand{\rp}{\right)}
\newcommand{\aq}{\alpha_s\left( Q^2 \right)}

\newcommand{\nf}{n_f)}
\newcommand{\nn}{\nonumber}

\newcommand{\dy}{\ttt{dy}}
\newcommand{\dlnlnQ}{\ttt{dlnlnQ}}

\newcommand{\GeV}{\;\mathrm{GeV}}
\newcommand{\TeV}{\;\mathrm{TeV}}
\newcommand{\as}{\alpha_s}

\newcommand{\eg}{e.g.\ }
\newcommand{\ie}{i.e.\ }
\newcommand{\cf}{cf.\ }
\newcommand{\MSbar}{\overline{\mathrm{MS}}}
\newcommand{\hoppet}{\textsc{hoppet}\xspace}
\newcommand{\ttt}[1]{\texttt{#1}}
\newcommand{\order}[1]{{\cal O}\left(#1\right)}
\newcommand{\fn}{\scriptsize}

\newcommand{\AllDGLAP}{Botje,Schoeffel:1998tz,Pegasus,Pascaud:2001bi,Weinzierl:2002mv,coriano,GuzziThesis,nnpdf,Kosower:1997hg,Ratcliffe:2000kp}

\usepackage{color}
\definecolor{comment}{rgb}{0,0.3,0}
\definecolor{identifier}{rgb}{0.0,0,0.3}

\usepackage{listings}
\title{A Higher Order Perturbative Parton Evolution Toolkit \\
(\hoppet)
}
\author{G.~P. Salam and J.~Rojo\\[3pt]
  LPTHE,   \\UPMC -- Univ.\ Paris 6,\\
  Universit\'e Paris Diderot -- Paris 7,\\
  CNRS UMR 7589,\\ 75252 Paris cedex 05, France} 
\date{}

\begin{document}

\maketitle 

\begin{abstract}
  This document describes a Fortran~95 package for carrying out DGLAP
  evolution and other common manipulations of parton distribution
  functions (PDFs). The PDFs are represented on a grid in $x$-space so
  as to avoid limitations on the functional form of input
  distributions.  Good speed and accuracy are obtained through the
  representation of splitting functions in terms of their convolution
  with a set of piecewise polynomial basis functions, and Runge-Kutta
  techniques are used for the evolution in $Q$.  Unpolarised evolution
  is provided to NNLO, including heavy-quark thresholds in the
  $\MSbar$ scheme, and longitudinally polarised evolution to NLO. The
  code is structured so as to provide simple access to the objects
  representing splitting functions and PDFs, making it possible for a
  user to extend the facilities already provided.
  A streamlined interface is also available, facilitating use of the
  evolution part of the code from F77 and C/C++.  \smallskip
\end{abstract}

\newpage

\noindent {\Large \textbf{Program Summary}}\\

\noindent {\em Title of program\/}: \hoppet \\[2mm]
{\em Version\/}: 1.1 \\[2mm]
{\em Catalogue identifier\/}: \\[2mm]
{\em Program obtainable from\/}:
\url{http://projects.hepforge.org/hoppet/}
\\[2mm]
{\em Distribution format\/}: compressed tar file \\[2mm]
{\em E-mail\/}: {\tt salam@lpthe.jussieu.fr}, 
{\tt rojo@lpthe.jussieu.fr} \\[2mm]
{\em License\/}: GNU Public License \\[2mm]
{\em Computers\/}: all \\[2mm]
{\em Operating systems\/}: all \\[2mm]
{\em Program language\/}: Fortran~95 \\[2mm]
{\em Memory required to execute\/}:  $\lesssim$ 10 MB \\[2mm]
{\em Other programs called\/}: none \\[2mm]
{\em External files needed\/}: none \\[2mm]
{\em Number of bytes in distributed program, including test data
  etc.\/}: $\sim 270$~kB\\[2mm]
\\[2mm]
{\em Nature of the physical problem\/}: Solution of the DGLAP
evolution equations up to NNLO (NLO) for unpolarised (longitudinally
polarised) PDFs, and provision of tools to facilitate manipulation
(convolutions, etc.) of PDFs with user-defined coefficient and
splitting functions.
\\[2mm]
{\em Method of solution\/}:
representation of PDFs on a grid in $x$, adaptive integration of
splitting functions to reduce them to a discretised
form, obtaining fast
convolutions that are equivalent to integration with an interpolated
form of the PDFs; Runge Kutta solution of the $Q$ evolution, and
its caching so as to speed up repeated evolution with different
initial conditions.
\\[2mm]
{\em Restrictions on complexity of the problem\/}: PDFs should be
smooth on the scale of the discretisation in $x$.
\\[2mm]
{\em Typical running time\/}: a few seconds for initialisation, then
$\sim 10$~ms for creating a tabulation with a relative accuracy of
$10^{-4}$ from a new initial condition (on a 3.4~GHz Pentium~IV
processor). Further details in Sect.~\ref{sec:Timing}.

\newpage
\tableofcontents

%======================================================================
%======================================================================
\section{Introduction}
\label{sec:intro}

There has been considerable discussion over the past years
(\eg~\cite{\AllDGLAP}) of numerical solutions of the 
Dokshitzer-Gribov-Lipatov-Altarelli-Parisi (DGLAP)
 equation~\cite{DGLAP} for the Quantum Chromodynamics
(QCD)
evolution  of parton distribution functions (PDFs).

The DGLAP equation
\cite{DGLAP} is a renormalisation group equation for the quantity
$q_i(x,Q^2)$, the density of partons of type (or flavour) $i$ carrying
a fraction $x$ of the longitudinal momentum of a hadron, when resolved
at a scale $Q$. It is one of the fundamental equations of perturbative
QCD, being central to all theoretical
predictions for hadron-hadron and lepton-hadron colliders.

Technically, it is a matrix integro-differential equation,
\begin{equation}
  \label{eq:dglap}
  \frac{\partial q_i(x,Q^2)}{\partial \ln Q^2} = \frac{\aq}{2\pi}
 \int_x^1 \frac{dz}{z}
  P_{ij}\lp z,\aq\rp q_j\lp \frac{x}{z},Q^2\rp\,,
\end{equation}
whose kernel elements $P_{ij}(z,Q^2)$ are known as splitting
functions, since they describe the splitting of a parton of kind $j$
into a parton of kind $i$ carrying a fraction $z$ of the longitudinal
momentum of $j$. The parton densities themselves $q_i(x,Q^2)$ are
essentially non-perturbative, since they depend on physics at hadronic
mass scales $\lesssim 1 \GeV$, where the QCD coupling is large. On the
other hand the splitting functions are given by a perturbative
expansion in the QCD coupling $\as(Q^2)$. Thus given partial
experimental information on the parton densities\footnote{Of course it
  is not the parton densities, but rather structure functions, which
  can be derived from them perturbatively, that are measured
  experimentally.} %
--- for example over a limited range of $Q$, or for only a subset of
parton flavours --- the DGLAP equations can be used to reconstruct the
parton densities over the full range of $Q$ and for all flavours.

The pivotal role played by the DGLAP equation has motivated a
considerable body of literature discussing its numerical solution
\cite{\AllDGLAP}. There exist two main classes of approaches: those that
solve the equation directly in $x$-space and those that solve it for
Mellin transforms of the
parton densities, defined as
\be
q_{N}\lp N,Q^2\rp = \int_0^1 dx x^N q_i(x,Q^2) \ ,
\ee 
 and subsequently invert the transform back to
$x$-space.
 Recently, a novel approach has been proposed which combines
advantages of the $N-$space and $x-$space methods \cite{nnpdf}.
$N-$space based methods are of interest because the Mellin transform
converts the convolution of eq.~(\ref{eq:dglap}) into a multiplication,
resulting in a continuum of independent matrix differential (rather
than integro-differential) equations, one for each value of $N$,
making the evolution more efficient numerically.

The drawback of the Mellin method is that one needs to know the Mellin
transforms of both the splitting functions and the initial conditions.
There can also be subtleties associated with the inverse Mellin
transform.
The $x$-space method is in contrast more flexible, since the inputs
are only required in $x$-space; however it is generally considered to
less efficient numerically, because of the need to carry out the
convolution in eq.~(\ref{eq:dglap}).

To understand the question of efficiencies one should analyse the
number of operations needed to carry out the evolution. Assuming that
one needs to establish the results of the evolution at $N_x$ values of
$x$, and $N_Q$ values of $Q$, one essentially needs $\order{N_x^2
  N_Q}$ operations with an $x$-space method, where the $N_x^2$ factor
comes from the convolutions. In the Mellin-space method, one needs
$\order{N_x N_Q M}$ operations, where $M$ is the number of points used
for Mellin inversion. One source of drawback of the $x$-space method
is that, nearly always, $N_x \sim \ln 1/x_{\min}$ and so the method
scales as the square of $\ln 1/x_{\min}$, where the Mellin method is
linear (and $M$ can be kept roughly independent of $x_{\min}$). 

The other issue relates to how one goes to higher numerical
integration and interpolation orders.
In $x$-space methods one tends to choose $x$ values that are uniformly
distributed (be it in $\ln 1/x$ or some other more complex function)
--- this limits one to higher-order extensions of the Trapezium and
Simpson-rule type integrations, whose order in general is $n_p-1$
where $n_p$ is the number of points used for the integration. The
precision of the integration is given by $\lp\delta x\rp^{n_p}$ where
$\delta x$ is the grid spacing. Higher $n_p$ improves the accuracy,
but typically $n_p$ can not be taken too large because of large
cancellations between weights that arise for large $n_p$.
In the Mellin method one is free to position the $M$ points as one
likes, and one can then use Gaussian type
integration~\cite{Weinzierl:2002mv,Pegasus,Kosower:1997hg}; using $n_p$
points one manages to get a numerical order $2n_p-1$, \ie accuracy
$\lp \delta x\rp^{2n_p}$, and furthermore the integration weights do not suffer
from cancellations at large $n_p$, allowing one to increase
$n_p$, and thus the accuracy, quite considerably.

Despite it being more difficult to obtain high accuracy with $x$-space
methods, their greater flexibility means that they are widespread,
serving as the basis of the well-known  QCDNUM program \cite{Botje},
and  used also by the CTEQ~\cite{CTEQ} and MRST/MRSW~\cite{MRST}
global fitting
collaborations. Higher-order methods in $x$-space have been developed
in \cite{Schoeffel:1998tz,Pascaud:2001bi,coriano,GuzziThesis,DisResum}, and
more recently have been incorporated also in QCDNUM.

\hoppet, the program presented here, uses higher-order methods both for
the $x$-integrations and $Q$ evolution. It combines this with multiple
grids in $x$-space: a high-density grid at large $x$ where it is
hardest to obtain good accuracy, and coarser grids at smaller $x$
where the smoothness of the PDFs facilitates the integrations. One of
the other crucial features of the program is that it pre-calculates as
much information as possible, so as to reduce the 
evolution of a new PDF initial condition to a modest set of addition
and multiplication operations. Additionally, the program provides
access to a range of low and medium-level operations on PDFs which
should allow a user to extend the facilities already provided.

The functionality described in this article has been present in
\hoppet's predecessors for several years (they were available on
request), but had
never been documented. 
Those predecessors have  been used in a number of different
contexts, like resummation of event shapes
in DIS \cite{DisResum}, automated resummation
of event shapes \cite{caesar}, studies of
resummation in the small-$x$ limit \cite{Ciafaloni:2003rd}, 
and in a posteriori inclusion
of PDFs in NLO final-state calculations \cite{APPL,Banfi:2007gu}, as well
as  used for benchmark
comparisons with Pegasus~\cite{Pegasus} in \cite{Benchmarks}.
Since the code had
not hitherto been released in a documented form, 
it is the authors' hope that availability of
this documentation may make the package somewhat more useful.

This manual is structured as follows: Sect.~\ref{sec:pqcd}
briefly summarises the perturbative QCD ingredients contained in
\hoppet, while Sect.~\ref{tricks} describes the numerical techniques
used to solve the DGLAP equation. 
Sects.~\ref{sec:singleflav}-\ref{sec:tabulated-pdfs} present in detail the capabilities
of the \hoppet package with its general F95 interface, 
with emphasis on those aspects that
can be adapted by a user to tailor it to their own needs. 
Sect.~\ref{sec:vanilla} describes a streamlined interface
to \hoppet which embodies its essential capabilities in
a simple interface available a variety of programming languages: 
F77 and C/C++.
Finally, Sect.~\ref{sec:benchmarks} presents a detailed 
quantitative study of
the performance of \hoppet, and in the final section we conclude.
A set of appendices contain various example programs,
both for the general and the streamlined interfaces, a reference
guide with the most important \hoppet modules,
details on technical aspects and a set of useful tips on
Fortran 95.

A reader 
whose interest is to use \hoppet to perform
fast and efficient evolution of PDFs 
may wish to skip 
Sects.~\ref{sec:singleflav} to \ref{sec:tabulated-pdfs}
and move directly to
 Sect.~\ref{sec:vanilla}, which describes the 
user-friendlier streamlined interface,
accessible from F95, F77 and C/C++, and which contains the essential
functionalities of \hoppet. He/she is also encouraged to go through the
various example programs which contain detailed 
descriptions and explanations. 
On the other hand, a reader interested in the
more flexible and general 
functionalities of \hoppet, should also consult
Sects.~\ref{sec:singleflav} to \ref{sec:tabulated-pdfs}. 

Note 
that throughout this documentation, a PDF refers always to a momentum
density rather than a parton density, that is, when we refer to
a gluon, we mean $xg(x)$ rather than $g(x)$, the same convention
as used in the LHAPDF PDF library \cite{LHAPDF}.

%----------------------------------------------------------------------

\section{Perturbative evolution in QCD}
\label{sec:pqcd}
First of all we set up the notation and
conventions that are used through \hoppet. The DGLAP
equation for a non-singlet parton distribution reads
\begin{equation}
  \label{eq:dglap-ns}
  \frac{\partial q(x,Q^2)}{\partial \ln Q^2} = 
\frac{\aq}{2\pi}\int_x^1 \frac{dz}{z}
  P(z,\aq) q\lp \frac{x}{z},Q^2\rp \equiv 
\frac{\aq}{2\pi}  P(x,\aq) \otimes q\lp x,Q^2\rp \ .
\end{equation}
Note that the related variable $t\equiv \ln Q^2$ is also used
through \hoppet.
The splitting functions in  eq.~(\ref{eq:dglap-ns})
are known up to NNLO in the 
unpolarised case \cite{NNLO-NS,NNLO-singlet}:
\begin{equation}
  \label{eq:dpdf}
   P(z,\aq)=P^{(0)}(z)+\frac{\aq}{2\pi}P^{(1)}(z)+
\lp \frac{\aq}{2\pi} \rp^2 P^{(2)}(z) \ ,
\end{equation}
and up to NLO in the polarised case.
The generalisation to the singlet case is straightforward, as it
is the generalisation of eq.~(\ref{eq:dglap-ns}) to the case
of time-like evolution\footnote{
The general structure of the relation between space-like
and time-like evolution and splitting functions
 has been investigated in \cite{Dokshitzer:2005bf,Mitov:2006ic,Basso:2006nk,Dokshitzer:2006nm,Beccaria:2007bb}.}, 
relevant for example for fragmentation function analysis,
where partial NNLO results
are also available \cite{Mitov:2006ic}.

As with the splitting functions, all perturbative quantities in
\hoppet are defined to be a coefficient of $\as/2\pi$. The one
exception is the $\beta$-function coefficients of the running
coupling equation:
\begin{equation}
  \label{eq:as-ev}
  \frac{d\as}{d\ln Q^2} = \beta\lp \aq\rp = -\as (\beta_0\as +
  \beta_1\as^2 + 
  \beta_2\as^3) \ .
\end{equation}

The evolution of the strong coupling and the parton distributions can
be performed in both the fixed flavour-number scheme (FFNS) and the 
variable flavour-number scheme (VFNS). In the VFNS case we 
need the matching conditions between the effective
theories with $n_f$ and $n_{f}+1$ light flavours for both the strong 
coupling $\aq$ and the parton distributions at the heavy quark
mass threshold $m_h^2$.

These matching conditions for the parton distributions
receive non-trivial contributions beginning at NNLO.
For light quarks $q_{l,i}$ of flavour $i$ 
(quarks that are considered massless
below the heavy quark mass threshold $m_h^2$) the matching between
their values in the $n_f$ and
$n_f+1$ effective theories reads:
\be
\label{eq:lp-nf1}
  q_{l,i}^{\,(n_f+1)}(x,m_h^2) \: = \:  q_{l,i}^{\,(\nf}(x,m_h^2) +
\lp \frac{\alpha_s(m_h^2)}{2\pi} \rp^2
   A^{\rm ns,(2)}_{qq,h}(x) \otimes
  q_{l,i}^{\, (\nf}(x,m_h^2) \ ,
\ee
where  $i = 1,\ldots n_f$, while for the gluon
distribution and the heavy quark PDF $q_h$ one has a coupled matching 
condition:
\bea
\label{eq:hp-nf1}
  g^{(n_f+1)}(x,m_h^2) \:\:\: &\! = \!\! &
    g^{\, (\nf}(x,m_h^2) 
\\ &+& \lp \frac{\alpha_s(m_h^2)}{2\pi} \rp^2 \Big[
    A_{\rm gq,h}^{S,(2)}(x) \otimes \Sigma^{(\nf}(x,m_h^2) +
    A_{\rm gg,h}^{S,(2)}(x) \otimes g^{(\nf}(x,m_h^2) \Big ]\ ,
  \nn \\[0.5mm] 
  (q_h+\bar{q}_{h})^{(n_f+1)}(x,m_h^2) \! &\! =\!\! &
   \lp \frac{\alpha_s(m_h^2)}{2\pi} \rp^2 \Big [
    \tilde{A}_{\rm hq}^{S,(2)}(x)\otimes \Sigma^{(\nf}(x,m_h^2) 
    + \tilde{A}_{\rm hg}^{S,(2)}(x)\otimes g^{(\nf}(x,m_h^2) \Big  ] \ ,
  \quad \nonumber
\eea
with $q_h=\bar{q}_h$, and the singlet PDF $\Sigma(x,Q^2)$ is defined
in Table \ref{eq:diag_split}.
The NNLO matching coefficients were
computed in \cite{NNLO-MTM}\footnote{The authors are thanked 
for the code corresponding to the
calculation.}.
Notice that the above conditions will lead to small discontinuities
of the PDFs in its evolution in $Q^2$, 
which are cancelled by similar matching terms
in the coefficient functions resulting in continuous physical
observables. In particular, the heavy quark PDFs start from a non-zero
value at threshold at NNLO, which sometimes can even be
negative.

The corresponding NNLO relation for the matching of the coupling constant 
at the heavy quark threshold $m^2_h$ is given by
\be
\label{eq:as-nf1}
  \as^{\, (n_f+1)}(m_h^2) \: = \:
  \as^{\, (\nf} (m_h^2) +   C_2 \lp \frac{\as^{\, (\nf} (m_h^2)}{2\pi} \rp^3
   \:\: ,
\ee
where the matching coefficient $C_2$ was computed in \cite{Chetyrkin:1997sg}.

The default basis for the PDFs, called the \ttt{human} 
representation in \hoppet, is such that 
 the entries in an array
\ttt{pdf(-6:6)} of PDFs correspond to:
\bea 
\bar t={-6} \ ,  \bar b={-5} \ ,  \bar c={-4}
\ , \nn   \bar s&=&{-3} \ , \nn  \bar u={-2} \ , \nn
 \bar d={-1} \ , \\  g&=&{0} \ , \\ \nn   d={1} \ , \nn  u={2} 
\ , \nn  
s={3} \ , \nn   c&=&{4} \ , \nn b={5} \ , \nn  t={6} \ . \nn 
\eea
 This representation is the
same as that used through the \ttt{LHAPDF} library \cite{LHAPDF}. 
However, this representation leads
to a complicated form of the evolution equations.
The splitting matrix can be simplified considerably (made diagonal
except for a $2\times2$ singlet block) by switching to a different
flavour representation, which is named
the \ttt{evln} representation, for the PDF set, as explained in detail in
\cite{vanNeerven:1999ca,vanNeerven:2000uj}. This representation
is described in Table \ref{eq:diag_split}.

In the {\tt evln} basis, 
the gluon evolves coupled to the singlet  PDF $\Sigma$,
and all non-singlet PDFs evolve independently.
Notice that the representations of the PDFs
are preserved under linear operations, so in particular
they are preserved under DGLAP evolution.
The conversion from the \ttt{human} to the \ttt{evln}
representations of PDFs requires that the number of
active quark flavours $n_f$ is specified by the user, as described in
Sect. \ref{sec:evln-rep}.

\begin{table}
\begin{center}
\begin{tabular}{|r | c | l |}
\hline
     i & \mbox{name} & $q_i$ \\ \hline
     $ -6\ldots-(n_f+1)$ & $q_i$ & $q_i$\\
     $-n_f\ldots -2$ & $q_{\mathrm{NS},i}^{-}$ & 
$(q_i -  {\bar q}_i) - (q_1 - {\bar q}_1)$\\
      -1           & $q_{\mathrm{NS}}^{V}$ & 
$\sum_{j=1}^{n_f} (q_j -  {\bar q}_j)$\\
       0           & g & \textrm{gluon} \\
       1           & $\Sigma$ & $\sum_{j=1}^{n_f} (q_j +  {\bar q}_j)$\\
     $2\ldots n_f$ & $q_{\mathrm{NS},i}^{+}$ &
$ (q_i +  {\bar q}_i) - (q_1 + {\bar q}_1)$\\
      $(n_f+1)\ldots6$ & $q_i$ & $q_i$ \\
\hline
\end{tabular}
\caption{}{\label{eq:diag_split} The evolution representation 
(called \ttt{evln} in \hoppet)
of PDFs with $n_f$ active quark flavours
in terms of the \ttt{human} representation.}  
\end{center}
\end{table}

In \hoppet unpolarised DGLAP evolution is available up to NNLO
in the $\MSbar$ scheme, while for the DIS scheme
only evolution up to NLO is available. For polarised evolution only
the $\MSbar$ scheme is available. The variable \ttt{factscheme}
takes different values for each factorisation scheme:
\begin{center}
  \begin{tabular}{|c|l|}\hline
    \ttt{factscheme} & Evolution\\[2pt]\hline
    1 & unpolarised $\MSbar$ scheme\\[2pt]\hline
    2 & unpolarised DIS scheme\\[2pt]\hline
    3 & polarised $\MSbar$ scheme\\\hline
  \end{tabular}
\end{center}
Note that mass thresholds are currently
missing in the DIS scheme.

%-----------------------------------------------------------------
\section{Numerical techniques}
\label{tricks}
We briefly introduce now the numerical techniques used to perform
parton evolution: the discretisation of PDFs and
their convolutions with splitting functions on a grid in $x$,
and the subsequent DGLAP evolution in $Q^2$. 

The first aspect that we discuss is how to represent PDFs
and associated convolutions in terms of an interpolating grid
in $x-$space.
Note that this technique can be applied to any quantity
that appears through convolutions, so it is not restricted
to parton distributions. Then we will describe how
to obtain the solution of the DGLAP evolution equations.

%......................................................................
\subsection{Higher order matrix representation}
\label{sec:highord}

Given a set of $N_x$ grid points $y_\alpha=\ln 1/x_{\alpha}$, 
labelled by an
index $\alpha$ and
(for later convenience) a uniform grid spacing, $y_\alpha = \alpha \delta
y$, one can approximate a parton distribution function 
$xq \lp x,t \rp$ by
interpolating the PDF at the grid points,\footnote{From the 
numerical point of view, it is
advantageous to interpolate $xq$ rather than $q$ itself because
the is in general smoother.}
\begin{equation}
  xq(y=\ln 1/x,t) = \sum_\alpha w_\alpha(y) q_\alpha(t)\,,
\end{equation}
where $w_\alpha(y)$ are the interpolation weights, 
we have defined
\begin{equation}
 q_{\alpha}(t) \equiv
x_{\alpha}q(y_\alpha,t) \ ,
\end{equation}
and the sum over $\alpha$ runs over $n+1$ points in the
vicinity of $y$ for an interpolation order $n$. Note that Greek
indices represent the $y$ dimension, while  Roman indices
are  used to represent the flavour dimension.

The convolution of a single-flavour PDF with a splitting function
$P(z,t)$ can be written as
\begin{equation}
  (P \otimes q)(y,t) = \sum_{\alpha} \omega_{\alpha}(y)
 (P \otimes q)_\alpha(t) \ ,
\end{equation}
where we have replaced the convolution by its grid representation,
\begin{equation}
  (P \otimes q)_\alpha(t) = \sum_{\beta} P_{\alpha\beta}(t) \, q_\beta(t)\,,
\end{equation}
where $\beta$ runs over $\mathcal{O}\lp N_x \rp$ points of the grid and we have
defined
\begin{equation}
  \label{eq:Palphabeta}
  P_{\alpha\beta}(t) = \int_{e^{-y_{\alpha}}}^1 dz
 \,P(z,t)\, w_\beta(y_\alpha + \ln z)\  .
\end{equation}

Note that the piecewise  interpolating polynomials 
that we use are not smooth at the
grid points. The reason for this is that this type of interpolation
strategy is much more suited for the numerical integration required
in convolutions than other strategies which impose continuity
on the interpolating polynomials, like for example splines, which
are in general slower and less accurate for the same number
of grid points.

A virtue
 of having a uniform grid in $y = \ln 1/x$ is that the
interpolation functions can be arranged to have a structure
$w_\beta(y) = w_{\beta}(y - y_\beta)$, so that $P_{\alpha\beta}$ just depends
on $\alpha - \beta$, and can be rewritten $\mathcal{P}_{\alpha - \beta}$. A
slight subtlety arises at large $x$, where if one writes $w_\beta(y) =
w_{\beta}(y - y_\beta)$ one is effectively assuming an interpolation that uses
identically zero interpolation points for $x\ge 1$~\cite{coriano,
Ratcliffe:2000kp},
even though $xq(x)$ is not formally defined for $x > 1$. In practice
this is often not too important (because PDFs drop rapidly towards
$x=1$), but we shall include two options: one that uses effective
zero-points beyond $x=1$ and one that ensures that the interpolation
is based only on the physically valid domain of the PDFs.

\begin{figure}
  \centering
  \includegraphics[width=0.72\textwidth]{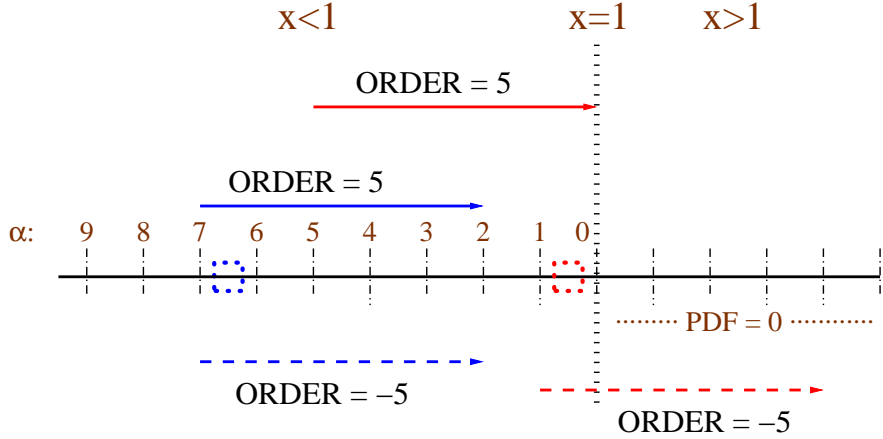}%
  \caption{The different strategies (+ve and $-$ve order) for
    interpolation of the grid near $x=1$. The dotted (blue, red) boxes
    indicate two regions in which we illustrate the interpolation of
    the PDF, while the (blue, red) lines with arrows indicate the
    corresponding range of grid points on which the interpolation is
    based.}
  \label{fig:interpolation-orders}
\end{figure}

These two choices are represented in
Fig.~\ref{fig:interpolation-orders}.  For an interpolation of order
$n$ (that is, which uses information from $n+1$ grid points), the
option of using only points with $x\le 1$ is denoted by $\tt order=n$.
This has the consequence that for $\beta \le n$ we cannot write
$P_{\alpha\beta} = \mathcal{P}_{\alpha-\beta}$, and so must explicitly
store $\order{N_x n}$ distinct $P_{\alpha\beta}$ entries.
The option of using artificial (zero-valued) points at $x>1$ is denoted by $\tt
order=-n$, and does allow us to write $P_{\alpha\beta} =
\mathcal{P}_{\alpha-\beta}$ (thus we store only $\order{N_x}$
entries), with $\mathcal{P}_{\alpha-\beta} = 0$ for $\alpha < \beta$.

%......................................................................
\subsection{Evolution operators}
\label{sec:evop}
The DGLAP evolution equation, eq.~(\ref{eq:dglap-ns}), is easily 
approximated in terms of its grid representation by
\begin{equation}
\label{eq:evol}
  \frac{\partial q_{\alpha}(t)}{\partial t}  = \frac{\alpha_s(t)}{2\pi}
\sum_{\beta} P_{\alpha\beta}(t) q_\beta(t) \ ,
\end{equation}
where for a general value of $\alpha$ the sum over $\beta$ extends over
$\mathcal{O}\lp \alpha + |\ttt{order}| \rp$ 
points of the grid.
Introducing $M_{\alpha\beta}(t_0) = \delta_{\alpha\beta}$ as initial
condition at some initial scale $t_0$, one can
alternatively solve
\begin{equation}
\label{eq:evolop}
  \frac{\partial M_{\alpha\beta}(t) }{\partial t} = \frac{\alpha_s(t)}{2\pi}
\sum_{\gamma} P_{\alpha\gamma}(t)
  M_{\gamma\beta}(t)\,.
\end{equation}
Then the evolved parton distribution at the grid points is
given by
\begin{equation}
  q_{\alpha}(t)  = \sum_{\beta} M_{\alpha\beta}(t) q_\beta(t_0)\,.
\end{equation}
We refer to $M_{\alpha\beta}(t)$ as the evolution operator. From 
a practical point of view, we will solve eqs.~(\ref{eq:evol})
and ~(\ref{eq:evolop})
with higher order iterative Runge-Kutta methods, as described in
Sect. \ref{sec:direct-evolution}.

A further simplification occurs 
if one can rewrite the splitting functions $P$ as
a translation invariant object. Using the properties
discussed in the previous subsection, 
we can rewrite $P_{\alpha\beta} = \cal P_{\alpha-\beta}$,
and then similarly one can rewrite $M_{\alpha\beta} = \cal
M_{\alpha-\beta}$, and it is as simple to determine
$M_{\alpha\beta}(t)$ as it is to determine the evolution of a single
vector $q_\alpha$, i.e.\ one just evolves a single column, $\beta =
0$, of $M_{\alpha\beta}(t)$.

%----------------------------------------------------------------------
\section{Single-flavour grids and convolutions}
\label{sec:singleflav}

\hoppet is written in Fortran~95 (F95). This has considerable
advantages compared to F77, as will be seen in the discussion of the
program, though it does lack a number of fully object-oriented
features and this sometimes restricts the scope for expressiveness.
Fortran~95 perhaps not being the best known language in the
high-energy physics community, occasionally some indications will be give to
help the reader with less-known language constructs, with further
information in Appendix~\ref{sec:f95appendix}.

All routines described in this section need access to the
\texttt{convolution} module, which can either be obtained directly by
adding a 
\begin{lstlisting}
  use convolution
\end{lstlisting}
statement at the beginning of the relevant subprogram (before
any \ttt{implicit none} or similar declarations). Alternatively, as
with the rest of the routines described in this documentation, it can
be accessed indirectly through the \ttt{hoppet\_v1} module
\begin{lstlisting}
  use hoppet_v1
\end{lstlisting}
Unless you are delving into the innards of  \hoppet, the latter is
more convenient since it provides access to everything you are likely
to need.

%......................................................................
\subsection{Grid definitions (\texttt{grid\_def})}
\label{sec:grid}

The grid (in $y$) is the central element of the PDF evolution.
Information concerning the grid is stored in a derived type
\texttt{grid\_def}:
% which could be initialised  for example as follows:
\begin{lstlisting}
  type(grid_def) :: grid
  call InitGridDef(grid,dy=0.1_dp,ymax=10.0_dp,order=5)
\end{lstlisting}
This initialises a grid between $x=1$ and down to $x =
e^{-\texttt{ymax}}$, with uniform grid spacing in $y = \ln 1/x$ of
\texttt{dy=0.1}, with a grid which  uses order 5 interpolation
with only $x\le 1$ points.
The user can modify this choice to better
suit his/her particular needs, as explained
in Sect.~\ref{sec:highord}.
One notes 
the use of keyword arguments --- the keywords are not
mandatory in this case, but have been included to improve the
legibility. Having defined a grid, 
the user need not worry about the details of
the grid representation.

In line with the convention set out in the Fortran~90 edition of
Numerical Recipes \cite{NRf90} we shall use \texttt{\_dp} to indicate
that numbers are in double precision, and \ttt{real(dp)} to declare
double precision variables. The integer parameter \ttt{dp} is defined
in the \texttt{module types} (and available indirectly through
\ttt{module hoppet\_v1}).

It is often useful to have multiple grids, with
coarser coverage at small $x$ and finer coverage at high $x$, to
improve the precision the convolution\footnote{
The reason that denser grids are required at large-$x$ is that
if a typical parton distributions goes as
\be
\lim_{x\to 1} q(x,Q^2) \sim (1-x)^m \ ,
\ee
then its logarithmic derivative with respect to $x$ is divergent,
\be
\lim_{x \to 1} \frac{\partial  \ln q(x,Q^2) }{\partial \ln x} = 
\lim_{x \to 1} \frac{-mx}{1-x} \to -\infty \ ,
\ee
and therefore to maintain the relative accuracy of the evolution, grids
with denser coverage at large-$x$ are required.} at large-$x$ without
introducing an unnecessarily large density of points
at small-$x$.
To support
this option, we can first define an array of sub-grids, and then use them to
initialise a combined grid as follows:
\begin{lstlisting}
  type(grid_def) :: grid, subgrids(3)

  ! define the various sub-grids
  call InitGridDef(subgrids(1),dy=0.30_dp, ymax=10.0_dp, order=5)
  call InitGridDef(subgrids(2),dy=0.10_dp, ymax= 2.0_dp, order=5)
  call InitGridDef(subgrids(3),dy=0.03333_dp, ymax= 0.6_dp, order=5) 
               ! Smaller dy at small ymax / large xmin 

  ! put them together into a single combined grid
  call InitGridDef(grid, subgrids, locked=.true.)
\end{lstlisting}
When combining them, the \ttt{locked=.true.} option has been specified,
which ensures that after any convolution, information from the finer
grids is propagated into the coarser ones. This places some
requirements on the grid spacings, notably that a coarse grid have a
spacing that is a multiple of that of the next finest grid. If the
requirements are not satisfied by the \ttt{subgrids} that have been
provided, then new similar, but more suitable subgrids are
automatically generated. 

Note that the two kinds of invocation of \ttt{InitGridDef} actually
correspond to different (overloaded) subroutines. The Fortran~95 compiler
automatically selects the correct one on the basis of the types of
arguments passed.

Though only grids that are uniform in $y$ have been implemented
(and the option of a simultaneous combination of them),
nearly all of the description that follows and all code outside the
\texttt{convolution} module are independent of this detail, the only
exception being certain statements about timings. Therefore were
there to be a strong motivation for an alternative, non-uniform grid,
it would suffice to modify the \texttt{convolution} module, while the
rest of the library (and its interfaces) would remain unchanged.

%......................................................................
\subsection[$x$-space functions]{$\boldsymbol{x}$-space functions}
\label{sec:xspc}

Normal $x$-space functions (such as PDFs) are
held in double precision arrays, which are to be allocated as follows
\begin{lstlisting}
  real(dp), pointer :: xgluon(:)
  call AllocGridQuant(grid,xgluon)
\end{lstlisting}
Note that for this to work, \texttt{xgluon(:)} should be a
\texttt{pointer}, and not just 
have the \texttt{allocatable} attribute. To
deallocate a grid quantity, one may safely use the F95
\texttt{deallocate} command.

Since \texttt{xgluon(:)} is just an array, it carries no information
about the \texttt{grid}. Therefore to set and access its value, one
must always provide the information about the \texttt{grid}. This is
not entirely satisfactory, and is one of the drawbacks of the use of
F95.
% as will be explained later on.

There are a number of ways of setting a grid quantity. Suppose 
for example  that we have
a function
\begin{lstlisting}
  function example_xgluon(y)
    use types                !! defines "dp" (double precision) kind
    implicit none
    real(dp), intent(in)  :: y
    real(dp) :: x    
    x = exp(-y)
    example_xgluon = 1.7_dp * x**(-0.1_dp) * (1-x)**5 !! returns xg(x)  
  end function example_xgluon
\end{lstlisting}
which
returns the gluon
momentum density $xg(x)$ (\cf Sect.~\ref{sec:highord}).
Then we can call
\begin{lstlisting}
  call InitGridQuant(grid,xgluon,example_xgluon)
\end{lstlisting}
to initialise \texttt{xgluon} with a representation of the return value
from the function \texttt{example\_xgluon}.
Alternative methods for initialising grid quantities
are described in Appendix \ref{sec:gridinit}.

To then access the gluon at a given value of $y = \ln 1/x$, one
proceeds as follows
\begin{lstlisting}
  real(dp) :: y, xgluon_at_y
  ...
  y = 5.0_dp
  xgluon_at_y = EvalGridQuant(grid,xgluon,y) !! again this returns xg(x)
\end{lstlisting}
Note that again we have to supply 
the \texttt{grid} argument to \ttt{EvalGridQuant}
because the \ttt{xgluon} array itself carries no information about the
grid (other than its size).

A less efficient, but perhaps more `object-oriented' way of accessing
the gluon is via the notation 
\begin{lstlisting}
  xgluon_at_y = xgluon .aty. (y.with.grid)
\end{lstlisting}
There also exists an \ttt{.atx.} operator for evaluating the PDF at a
given $x$ value.  Many of these procedures and operators are
overloaded so as to work with higher-dimensional arrays of grid
quantities, for example a multi-flavour
PDF array \texttt{pdf(:,:)}. The first index will
always correspond to the representation on the grid, while the second
index would here indicate the flavour.

Note that arithmetic operators all have higher precedence than
library-defined operators such as \texttt{.aty.}; accordingly some
ways of writing things are more efficient than others:
\begin{lstlisting}
  xgluon_at_y_times_2 = 2 * xgluon .aty. (y.with.grid)   ! very inefficient
  xgluon_at_y_times_2 = 2 * (xgluon .aty. (y.with.grid)) ! fairly efficient
  xgluon_at_y_times_2 = 2 * EvalGridQuant(grid,xgluon,y) ! most efficient
\end{lstlisting}
In the first case the whole of the array \texttt{xgluon} is multiplied
by 2, and then the result is evaluated at $y$, whereas in the second
and third
cases only the result of the gluon at $y$ is multiplied by 2.

%......................................................................
\subsection{Grid convolution operators}
\label{sec:conv}

While it is relatively straightforward internally to represent a
grid-quantity (e.g. a PDF) as an array, for convolution
operators  it is generally useful to have certain extra
information. Accordingly a derived type has been defined to hold a
convolution operator, and routines are provided for allocation and
initialisation of splitting functions.
The following example describes how the $gg$ LO splitting function
would be used to initialise the corresponding 
convolution operator:
\begin{lstlisting}
  type(grid_conv) :: xPgg
  call AllocGridConv(grid,xPgg)
  call InitGridConv(grid,xPgg, xPgg_func)
\end{lstlisting}
where the $P_{gg}$ splitting function is provided in the form of the
function \texttt{xPgg\_func}. Note that this function
must return $xP_{gg}(x)$:
\begin{lstlisting}
  ! returns various components of exp(-y) P_gg (exp(-y))
  real(dp) function xPgg_func(y)
    use types
    use convolution_communicator ! provides cc_piece, and cc_REAL,...
    use qcd                      ! provides CA, TR, nf, ...
    implicit none
    real(dp), intent(in) :: y
    real(dp)             :: x

    x = exp(-y); xPgg_func = zero
    if (cc_piece == cc_DELTA) then   ! Delta function term
       xPgg_func = (11*CA - 4*nf*TR)/6.0_dp
    else
       if (cc_piece == cc_REAL .or. cc_piece == cc_REALVIRT) & 
           & xPgg_func = 2*CA*(x/(one-x) + (one-x)/x + x*(one-x))
       if (cc_piece == cc_VIRT .or. cc_piece == cc_REALVIRT) & 
           & xPgg_func = xPgg_func - 2*CA*one/(one-x)
       xPgg_func = xPgg_func * x     ! remember to return x * Pgg
    end if
  end function xPgg_func
\end{lstlisting}
To address the issue that convolution operators can involve
plus-distributions and delta functions, the module
\texttt{convolution\_communicator} contains a variable
\texttt{cc\_piece} which indicates which part of the splitting
function is to be returned --- the real, virtual, real + virtual, or
$\delta$-function pieces. 

The initialisation of a \texttt{grid\_conv} object uses an adaptive
Gaussian integrator (a variant of CERNLIB's \texttt{dgauss}) to
calculate the convolution of the splitting function with trial weight
functions. The default accuracy for these integrations is $10^{-7}$.
It can be modified to value \texttt{eps} with the following subroutine
call
\begin{lstlisting}
  call SetDefaultConvolutionEps(eps)
\end{lstlisting}
which is to be made before creating the \ttt{grid\_def}
object.Alternatively, an optional \ttt{eps} argument can be
  included in the call to \ttt{InitGridDef} as follows:
\begin{lstlisting}
  type(grid_def) :: grid
  real(dp) :: eps
  [ ... set eps ... ]
  call InitGridDef(grid,dy=0.1_dp,ymax=10.0_dp,order=3,eps)
 \end{lstlisting}
Note that {\tt eps} is just
one of the parameters affecting the final accuracy of convolutions. In
practice (unless going to extremely high accuracies) the grid spacing
and interpolation scheme are more critical.

Having allocated and initialised a \texttt{xPgg} splitting function, we
can go on to use it. For example: 
\begin{lstlisting}
  real(dp), pointer :: xPgg_x_xgluon(:)
  ...
  call AllocGridQuant(grid,xPgg_x_xgluon) !! Allocate memory for result of convolution
  xPgg_x_xgluon = xPgg .conv. xgluon      !! Convolution of xPgg with xgluon
\end{lstlisting}  
Since the return value of \texttt{xPgg .conv.\ xgluon} is just an F95
array, one can also write more complex expressions. Supposing we had
defined also a \texttt{xPgq} splitting function and a singlet
quark distribution \ttt{xquark}, 
as well as $\texttt{as2pi} = \as/2\pi$,
then to first order in $\as$ we could write the gluon evolution
through a step \texttt{dt} in $\ln Q^2$ as
\begin{lstlisting}
  xgluon = xgluon + (as2pi*dt) * ((xPgg .conv. xgluon) + (xPgq .conv. xquark))
\end{lstlisting}
Note that like \texttt{.aty.}, \texttt{.conv.} has a low precedence,
so the use of brackets is important to ensure that the above
expressions are sensible. Alternatively, the issues of precedence can
be addressed by using \texttt{*} (also defined as convolution when it
appears between a splitting function and a PDF) instead of
\texttt{.conv.}:
\begin{lstlisting}
  xgluon = xgluon + (as2pi*dt) * (xPgg*xgluon + xPgq*xquark)
\end{lstlisting}
Note that, for brevity, from now on we will drop the explicit use of
$x$ in front of names PDF and convolution operator variables.

%......................................................................
\subsubsection{Other operations on \texttt{grid\_conv} objects}
\label{sec:other_grid_conv_ops}
It is marginally less transparent to manipulate \texttt{grid\_conv} types
than PDF distributions, but still fairly simple:
\begin{lstlisting}
  call AllocGridConv(grid,Pab)         ! Pab memory allocated 
  call InitGridConv(grid,Pab)          ! Pab = 0             (opt.alloc)
  call InitGridConv(Pab,Pcd[,factor])  ! Pab = Pcd [*factor] (opt.alloc)
  call InitGridConv(grid,Pab,function) ! Pab = function      (opt.alloc)

  call SetToZero(Pab)                  ! Pab = 0
  call Multiply (Pab,factor)           ! Pab = Pab * factor
  call AddWithCoeff(Pab,Pcd[,coeff])   ! Pab = Pab + Pcd [*coeff]
  call AddWithCoeff(Pab,function)      ! Pab = Pab + function

  call SetToConvolution(Pab,Pac,Pcb)   ! Pab = Pac.conv.Pcb  (opt.alloc)
  call SetToConvolution(P(:,:),Pa(:,:),Pb(:,:))            ! (opt.alloc)
                                       ! P(:,:) = matmul(Pa(:,:),Pb(:,:))
  call SetToCommutator(P(:,:),Pa(:,:),Pb(:,:))             ! (opt.alloc)
                                       ! P(:,:) = matmul(Pa(:,:),Pb(:,:))
                                       !         -matmul(Pb(:,:),Pa(:,:))    

  call Delete(Pab)                     ! Pab memory freed
\end{lstlisting}
Routines labelled ``\texttt{(opt.alloc.)}'' allocate the memory for
the \texttt{grid\_conv} object if the memory has not already been
allocated. (If it has already been allocated it is assumed to
correspond to the same grid as any other \texttt{grid\_conv} objects
in the same subroutine call). Some calls require that one specify the
grid definition being used (\texttt{grid}), because otherwise there is
no way for the subroutine to deduce which grid is being used.

If repeatedly creating a \texttt{grid\_conv} object for temporary use, it is
important to remember to \texttt{Delete} it afterwards, so as to avoid
memory leaks.

Nearly all the routines are partially overloaded so as to be able to
deal with one and two-dimensional arrays of \texttt{grid\_conv}
objects as well. The exceptions are those that initialise the
\texttt{grid\_conv} object from a function (arrays of functions do not
exist), as well as the convolution routines (for which the extension
to arrays might be considered non-obvious) and the commutation routine
which only has sense for matrices of \texttt{grid\_conv} objects.

%......................................................................
\subsubsection{Derived \texttt{grid\_conv} objects}
\label{sec:derived_grid_conv}

Sometimes it can be cumbersome to manipulate the \texttt{grid\_conv}
objects directly, for example when trying to create a
\texttt{grid\_conv} that represents not a fixed order splitting
function, but the resummed evolution from one scale to another. For
such situations the following approach can be used
\begin{lstlisting}
  real(dp), pointer :: probes(:,:)
  type(grid_conv)   :: Pqg, Pgq, Presult
  integer           :: i

  call GetDerivedProbes(grid,probes) ! get a set of 'probes'
  do i = 1, size(probes,dim=2)       ! carry out operations on each of the probes
    probes(:,i) = Pqg*(Pgq*probes(:,i)) - Pgq*(Pqg*probes(:,i))
  end do
  call AllocGridConv(grid,Presult)
  call SetDerivedConv(Presult,probes) ! Presult = [Pqg,Pgq]
\end{lstlisting}
Here \texttt{GetDerivedProbes} allocates and sets up an array of probe
parton distributions. Since a single-flavour parton distribution is a
one-dimensional array of \texttt{real(dp)}, the array of probes is a
two-dimensional array of \texttt{real(dp)}, the second dimension
corresponding to the index of the probe. One then carries out whatever
operations one wishes on each of the probes. Finally with the call to
\texttt{SetDerivedConv}, one can reconstruct a \texttt{grid\_conv}
object that corresponds to the set of operations just carried out

Some comments about memory allocation: the probes are automatically
allocated and deallocated; in contrast the call to
\texttt{SetDerivedConv(Presult,probes)} knows nothing about the grid,
so \texttt{Presult} must have been explicitly allocated for a specific
grid beforehand.

A note of caution: when one's grid is made of nested subgrids with the
locking option set to \texttt{.true.}, after a convolution of a
\texttt{grid\_def} object with a parton distribution, the coarser
grids for the parton distribution are supplemented with more accurate
information from the finer grids. When carrying out multiple
convolutions, this happens after each convolution.  There is no way to
emulate this with a single \texttt{grid\_def} object, and the locking
would actually confuse the reconstruction of resulting
\texttt{grid\_def} object. So when the user requests the probes,
locking is temporarily turned off globally and then reestablished
after the derived \texttt{grid\_object} has been constructed. Among
other things this means that acting with a derived
\texttt{grid\_object} will not be fully equivalent to carrying out the
individual operations separately. In particular the accuracy may be
slightly lower (whatever is lost due to the absence of intermediate
locking).

%======================================================================
%======================================================================
\section{Multi-flavour grids and convolutions}
\label{sec:dglapstructs}

The discussion in the previous section 
about how to represent functions and associated convolutions
in a general $x-$space grid holds for any kind of problem involving
convolutions, even if the examples were given in the context of DGLAP
evolution. In this section we shall examine the tools made available
specifically to address the DGLAP evolution problem.

%----------------------------------------------------------------------
\subsection{Full-flavour PDFs and flavour representations}
\label{sec:pdf-objects}

The routines described in this section
 are available from the \ttt{pdf\_general}
and \ttt{pdf\_representation} modules, or via the \ttt{hoppet\_v1}
general module.

Full flavour PDFs sets are just like single flavour PDFs except that they
have an extra dimension. They are represented by arrays, and if you
want \hoppet to deal with allocation for you, they should be pointer
arrays. One can allocate a single PDF (two dimensional
\texttt{real(dp)} array) or an array of PDFs (three-dimensional
\texttt{real(dp)} array)
\begin{lstlisting}
  real(dp), pointer :: PDF(:,:), PDFarray(:,:,:)
  call AllocPDF(grid,PDF)            ! allocates PDF(0:,-6:7)
  call AllocPDF(grid,PDFarray,0,10)  ! allocates PDFarray(0:,-6:7,0:10)
\end{lstlisting}
The first dimension corresponds to the grid in $y$; the second
dimension corresponds to the flavour index. Its lower bound is $-6$,
as one would expect. 

What takes a bit more getting used to is that its
upper bound is \textbf{7}. The reason is as follows: the flavour
information can be represented in different ways, for example each
flavour separately, or alternatively as singlet and non-singlet
combinations. In practice both are used inside the program and it is
useful for a PDF distribution to have information about the
representation, and this is stored in
\texttt{PDF(:,7)}\footnote{
In the current release of \hoppet, 
in particular, for PDFs in the \ttt{human} representation one
has \texttt{PDF(:,7)=0}, while for PDFs in the \ttt{evln}
representation, the information on the active number of flavours
is encoded as
\ttt{nf = (abs(q(2,7))+abs(q(3,7)))/(abs(q(0,7))+abs(q(1,7)))},
so that it is conserved under linear combinations.

However this feature might be modified in future
versions of the program.
}.

%......................................................................
\subsubsection{Human representation.}
\label{sec:human-rep}
When a PDF is allocated it is automatically labelled as being in the
\ttt{human} representation, described in
Sect. \ref{sec:pqcd}. Constants with names like
\ttt{iflv\_bbar}, \ttt{iflv\_g}, \ttt{iflv\_b}, are defined in
\ttt{module pdf\_representation}, to facilitate symbolic access to the
different flavours.

If you are creating a PDF as an automatic array (one whose bounds are
decided not by the allocation routine, but on the fly), for example in
a function that returns a PDF, then you should label it yourself as
being in the \ttt{human} representation, either with the
\ttt{LabelPdfAsHuman(pdf)} subroutine call, or by setting
\ttt{pdf(:,7)} to zero:
\begin{lstlisting}
module pdf_initial_condition
  use hoppet_v1; implicit none
contains
  function unpolarized_dummy_pdf(xvals) result(pdf)
    real(dp), intent(in) :: xvals(:)
    real(dp)             :: pdf(size(xvals),-6:7)

    ! clean method for labelling as PDF as being in the human representation
    call LabelPdfAsHuman(pdf)
    ! Alternatively, by setting everything to zero 
    ! (notably pdf(:,7)), the PDF representation
    ! is automatically set to be human
    pdf(:,:) = 0
    
    ! iflv_g is pre-defined integer parameter (=0) for symbolic ref. to gluon
    pdf(:,iflv_g) = 1.7_dp * xvals**(-0.1_dp) * (1-xvals)**5 ! Returns x*g(x)
    [... set other flavours here ...]
  end function unpolarized_dummy_pdf
end module pdf_initial_condition
\end{lstlisting}
The function has been placed in a module so as to provide an easy way
for a calling routine to have access to its interface (this is needed
for the dimension of \ttt{xvals} to be correctly passed).  Writing a
function such as that above is probably the easiest way of
initialising a PDF:
\begin{lstlisting}
  use hoppet_v1; use pdf_initial_condition; implicit none
  type(grid_def)    :: grid
  real(dp), pointer :: pdf(:,:)
  [...]
  call AllocPDF(grid,pdf)
  pdf = unpolarized_dummy_pdf(xValues(grid))
\end{lstlisting}
There exist a number of other options, which can be found by browsing
through \ttt{src/pdf\_general.f90}. Of these a sometimes handy one is
\begin{lstlisting}
  call AllocPDF(grid,pdf)
  call InitPDF_LHAPDF(grid, pdf, LHAsub, Q)
\end{lstlisting}
where \texttt{LHAsub} is the name of a subroutine 
with the same interface as LHAPDF's \ttt{evolvePDF} \cite{LHAPDF}:
\begin{lstlisting}
  subroutine LHAsub(x,Q,res)
    use types; implicit none
    real(dp), intent(in)  :: x,Q
    real(dp), intent(out) :: res(-6:6) ! on output contains flavours -6:6 at x,Q
    [...]                              ! Note that it should return momentum densities
  end subroutine LHAsub
\end{lstlisting}
Note that {\tt LHAsub} should return momentum densities, as
happens with the LHAPDF routines \cite{LHAPDF}.

Having initialised a PDF, to then extract it at a given $y$ value, one
can either examine a particular flavour using the methods described in
Sect.~\ref{sec:xspc}
\begin{lstlisting}
  real(dp) :: y, gluon_at_y
  gluon_at_y = pdf(:,iflv_g) .aty. (y.with.grid)
  ! OR
 gluon_at_y = EvalGridQuant(grid,pdf(:,iflv_g),y) 
\end{lstlisting}
or one can extract all flavours simultaneously
\begin{lstlisting}
  real(dp) :: pdf_at_y(-6:6)
  pdf_at_y = pdf(:,-6:6) .aty. (y.with.grid)
  ! OR
  pdf_at_y =  EvalGridQuant(grid,pdf(:,-6:6),y) 
\end{lstlisting}
with the latter being more efficient if one needs to extract all
flavours simultaneously. Note that here we have explicitly specified
the flavours, \ttt{-6:6}, that we want.\footnote{If instead we had
  said \ttt{pdf(:,:)} the result would have corresponded to a slice of
  flavours \ttt{-6:7}, \ie including an interpolation of the
  representation labelling information, which would be meaningless.}

%......................................................................
\subsubsection{Evolution representation} 
\label{sec:evln-rep}
For the purpose of carrying out convolutions, the \ttt{human}
representation is not very advantageous because the splitting matrix
in flavour space is quite complicated. Accordingly \hoppet uses a
different representation of the flavour internally when carrying out
convolution of splitting matrices with PDFs. For most purposes the
user need not be aware of this. The two exceptions are when a user
plans to create derived splitting matrices (being careless about the
flavour representation will lead to mistakes) or wishes to carry out
repeated convolutions for a fixed $n_f$ value (appropriate manual
changes of the flavour representation can speed things up).

The splitting matrix can be simplified considerably by switching to a different
flavour representation, as can be seen in Table \ref{eq:diag_split}. 
 When carrying out a convolution, the only non-diagonal part is the
block containing indices $0,1$. This representation is referred to as
the \ttt{evln} representation. Whereas the \ttt{human} representation
is $n_f$-independent the \ttt{evln} depends on $n_f$ through the
$\Sigma$ and $q_{\mathrm{NS}}^{V}$ entries and the fact that flavours
beyond $n_f$ are left in the human representation (since they are
inactive for evolution with $n_f$ flavours).

To take a PDF in the \ttt{human} representation and make a copy in an
\ttt{evln} representation, one uses the \ttt{CopyHumanPdfToEvln} routine
\begin{lstlisting}
  real(dp), pointer :: pdf_human(:,:), pdf_evln(:,:)
  integer           :: nf_lcl ! NB: nf would conflict with global variable
  [... setting up pdf_human, nf_lcl, etc. ...] 
  call AllocPDF(grid,pdf_evln)  ! or it might be an automatic array
  call CopyHumanPdfToEvln(nf_lcl, pdf_human, pdf_evln) ! From human to evolution representation
\end{lstlisting}
where one specifies the $n_f$ value for the \ttt{evln} representation.
One can go in the opposite direction with
\begin{lstlisting}
  call CopyEvlnPdfToHuman(nf_lcl, pdf_evln, pdf_human)
\end{lstlisting}
At any time one can check which is the representation of a given
PDF using the \ttt{GetPdfRep} function,
\begin{lstlisting}
  integer nf_rep
  real(dp), pointer :: pdf(:,:)

  [... set up pdf, ...]
  nf_rep = GetPdfRep(pdf)
\end{lstlisting}
which returns the number of active flavours if the PDF is in the
\ttt{evln} representation, or a negative integer if 
the PDF is in the
\ttt{human} representation.

%----------------------------------------------------------------------
\subsection{Splitting function matrices}
\label{sec:splitt-funct-matr}

Splitting function matrices and their actions on PDFs are defined in
   ~\ttt{module dglap\_objects} (accessible as usual from \ttt{module
  hoppet\_v1}). They have type \ttt{split\_mat}. Below we shall discuss
routines for creating specific predefined DGLAP splitting matrices,
but for now we consider a general splitting matrix.

The allocation of \ttt{split\_mat} objects,
\begin{lstlisting}
  type(split_mat) :: P
  integer         :: nf_lcl
  call AllocSplitMat(grid, P, nf_lcl)
\end{lstlisting}
is similar to that for \ttt{grid\_conv} objects. The crucial difference is
that one must supply a value for $n_f$, so that when the splitting
matrix acts on a PDF it knows which flavours are decoupled. From the
point of view of subsequent initialisation a \ttt{split\_mat} object
just consists of a set of splitting functions. 
If need be, they can be
initialised by hand, for example
\begin{lstlisting}
  call InitGridConv(grid,P%qq      , P_function_qq      )
  call InitGridConv(grid,P%qg      , P_function_qg      )
  call InitGridConv(grid,P%gq      , P_function_gq      )
  call InitGridConv(grid,P%gg      , P_function_gg      )
  call InitGridConv(grid,P%NS_plus , P_function_NS_plus )
  call InitGridConv(grid,P%NS_minus, P_function_NS_minus)
  call InitGridConv(grid,P%NS_V    , P_function_NS_V    )
\end{lstlisting}
We can then write
\begin{lstlisting}
  real(dp), pointer :: q(:,:), delta_q(:,:)
  [... allocations, etc. ...]
  delta_q = P .conv. q
  ! OR
  delta_q = P * q
\end{lstlisting}
and $\ttt{delta\_q}$ will have the following components
\begin{align}
  \label{eq:Pmat_on_q}
  \left(\!\!
    \begin{array}{c}
      \delta\Sigma\\
       \delta g
    \end{array}
  \!\!\right)
    \;&= \;
  \left(
    \begin{array}{cc}
      \ttt{P\%qq} & \ttt{P\%qg}\\
      \ttt{P\%gq} & \ttt{P\%gg}
    \end{array}
  \right) 
  \otimes
  \left(\!\!
    \begin{array}{c}
      \Sigma\\
       g
    \end{array}
    \!\!\right) 
  \nonumber\\[3pt]
  \delta q^+_{\mathrm{NS},i} \;&=\; \ttt{P\%NS\_plus} \otimes
  q^+_{\mathrm{NS},i}\\[3pt] 
  \delta q^-_{\mathrm{NS},i} \;&=\; \ttt{P\%NS\_minus} \otimes
  q^-_{\mathrm{NS},i}\nonumber \\[3pt]
  \delta q^V_{\mathrm{NS}} \;&=\; \ttt{P\%NS\_V} \otimes
  q^V_{\mathrm{NS}} \nonumber
\end{align}
We have written the result in terms of components in the \ttt{evln}
representation (and this is the representation used for the actual
convolutions). When a convolution with a PDF in \ttt{human}
representation is carried out, the program automatically copies the
PDF to the \ttt{evln} representation, carries out the convolution and
converts the result back to the \ttt{human} representation.
The cost of changing a representation is $\order{N_x}$, whereas the
convolution is $\order{N^2_x}$, so in principle the former is
negligible. In practice, especially when aiming for high speed at low
$N_x$, the change of representation can imply a significant cost. In
such cases, if multiple convolutions are to be carried out, it may be
advantageous to manually change into the appropriate \ttt{evln}
representation, carry out all the convolutions and then change back to
the \ttt{human} manually representation at the end, see
Sect.~\ref{sec:pdf-objects}.

As for \ttt{grid\_conv} objects, a variety of routines have been
implemented to help manipulate splitting matrices:
\begin{lstlisting}
  type(split_mat) :: PA, PB, PC
  real(dp) :: factor

  call InitSplitMat(PA,PB[,factor])   ! PA = PB [*factor]   (opt.alloc)

  call SetToZero(PA)                  ! PA = 0
  call Multiply(PA,factor)            ! PA = PA * factor               
  call AddWithCoeff(PA,PB[,factor])   ! PA = PA + PB [*factor]

  call SetToConvolution(PA,PB,PC)     ! PA = PB*PC          (opt.alloc)
  call SetToCommutator(PA,PB,PC)      ! PA = PB*PC-PC*PB    (opt.alloc)
  
  call Delete(split_mat)              ! PA's memory freed
\end{lstlisting}

%......................................................................
\subsubsection{Derived splitting matrices}
\label{sec:derived-split-matrices}

As with \ttt{grid\_conv} objects, \hoppet provides means to construct
a \ttt{split\_mat} object that corresponds to an arbitrary series of
\ttt{split\_mat} operations, as long as they all involve the same
value of $n_f$. One proceeds in a very similar way as
in Sect.~\ref{sec:derived_grid_conv},
\begin{lstlisting}
  real(dp), pointer :: probes(:,:,:)
  type(split_mat)  :: PA, PB, Pcomm
  integer           :: i

  [...set nf_lcl,...]
  call GetDerivedSplitMatProbes(grid,nf_lcl,probes) ! get the probes
  do i = 1, size(probes,dim=3)               ! carry out operations on each probe
    probes(:,:,i) = PA*(PB*probes(:,:,i)) - PB*(PA*probes(:,:,i))
  end do
  call AllocSplitMat(grid,Pcomm,nf_lcl)      ! provide nf info in initialisation
  call SetDerivedConv(Pcomm,probes)          ! Presult = [Pqg,Pgq]
\end{lstlisting}
Note that we need to provide the number of active quark flavours
to \ttt{GetDerivedSplitMatProbes}.
As in Sect.~\ref{sec:derived_grid_conv}, we first need to set up
some `probe' PDFs (note the extra dimension compared to earlier, since
we also have flavour information; the probe index always corresponds
to the last dimension); then we act on those probes; finally we
allocate the splitting matrix, and set its contents based on the
probes, which are then automatically deallocated.

%----------------------------------------------------------------------
\subsection{The DGLAP convolution components}
\label{sec:dglap_holder}

%......................................................................
\subsubsection{QCD constants}
\label{sec:qcd}

The splitting functions that we set up will depend on various QCD
constants ($n_f$, colour factors), so it is useful to here to
summarise how they are dealt with within the program.

The treatment of the QCD constants is \emph{not} object oriented.
There is a module (\ttt{qcd}) that provides access to commonly used
constants in QCD:
\begin{lstlisting}
  real(dp) :: ca, cf, tr, nf
  integer  :: nf_int
  
  real(dp) :: beta0, beta1, beta2
  [ ... ]
\end{lstlisting}
Note that \ttt{nf} is in double precision
 --- if you want the integer value of $n_f$, use \ttt{nf\_int}. 

To set the value of $n_f$, call
\begin{lstlisting}
  integer :: nf_lcl
  call qcd_SetNf(nf_lcl)  
\end{lstlisting}
where we have used the local 
variable \ttt{nf\_lcl} to avoid conflicting with
the \ttt{nf} variable provided by the \ttt{qcd} module. Whatever you
do, do not simply modify the value of the \ttt{nf} variable by hand
--- when you call \ttt{qcd\_SetNf} it adjusts a whole set of other
constants (\eg the $\beta$ function coefficients) appropriately.

There are situations in which it's of interest to vary the other
colour factors of QCD, for example, if these
colour factors are to be determined from a fit
to deep-inelastic scattering experimental data. For that purpose, use
\begin{lstlisting}
  real(dp) :: ca_lcl, cf_lcl, tr_lcl
  call qcd_SetGroup(ca_lcl, cf_lcl, tr_lcl)
\end{lstlisting}
Again all other constants in the \ttt{qcd} module will be adjusted. A
word of caution: the NNLO splitting functions actually depend on a
colour structure that goes beyond the usual $C_A$, $C_F$ and $T_R$,
namely $d_{abc}d^{abc}$, which in the present version of \hoppet
is hard-wired to its default QCD value.

%......................................................................
\subsubsection{DGLAP splitting matrices}
\label{sec:dglap-split}

The module \ttt{dglap\_objects} includes a number of routines for
providing access to the \ttt{split\_mat} objects corresponding to
DGLAP splitting functions
\begin{lstlisting}
  type(split_mat) :: P_LO, P_NLO, P_NNLO
  type(split_mat) :: Pp_LO, Pp_NLO        ! polarised

  ! MSbar unpolarised case
  call InitSplitMatLO  (grid, P_LO)
  call InitSplitMatNLO (grid, P_NLO)
  call InitSplitMatNNLO(grid, P_NNLO)

  ! the MSbar polarised case...
  call InitSplitMatPolLO (grid, Pp_LO)
  call InitSplitMatPolNLO(grid, Pp_NLO)
\end{lstlisting}
In each case the splitting function is set up for the $n_f$ and
colour-factor values that are current in the $\ttt{qcd}$ module, as
set with the \ttt{qcd\_SetNf} and \ttt{qcd\_SetGroup} subroutine calls. If one
subsequently resets the $n_f$ or colour factor values, the \ttt{split\_mat}
objects continue to correspond to the $n_f$ and colour factor values
for which they were initially calculated.
With the above subroutines for initialising DGLAP splitting functions,
the normalisation is as given in eq.~(\ref{eq:dpdf}).

In practice, because convolutions take a time $\order{N^2}$, where
$N$ is the number of points in the grid, whereas
additions and multiplications take a time $\order{N}$, in a program it
is more efficient to first sum the splitting matrices and then carry out the
convolution,
\begin{lstlisting}
  type(split_mat)   :: P_sum
  real(dp), pointer :: q(:,:), dq(:,:)
  [ ... ]
  call InitSplitMat(P_sum, P_LO)            ! P_sum = P_LO
  call AddWithCoeff(P_sum, P_NLO, as2pi)    ! P_sum = P_sum + as2pi * P_NLO
  dq = (as2pi * dt) * (P_sum .conv. q)      ! Step dt in evolution
  call Delete(P_sum)                        ! Memory freed
\end{lstlisting}
Note the use of brackets in the line setting \ttt{dq}: all scalar
factors are first multiplied together ($\order{1}$) so that we only
have one multiplication of a PDF ($\order{N_x}$). Note also that we have
chosen to include the \ttt{(as2pi * dt)} factor as multiplying the
pdf, rather than the other option of multiplying {P\_sum}, \ie 
\begin{lstlisting}
  call Multiply(P_sum, (as2pi * dt))
  dq =  P_sum .conv. q
\end{lstlisting}
The result would have been identical, but splitting matrices with
positive interpolation \ttt{order} essentially amount to an
$\order{7\times \ttt{order} \times N}$ sized array, whereas the PDF is
an $\order{13 N}$ sized array and the for high positive orders that
are sometimes used, it is cheaper to multiply the latter. 

The default for the NNLO splitting functions are the interpolated
expressions, which are very fast
to evaluate. Other possibilities, like
the exact splitting functions or a previous set of approximated NNLO
splitting functions which was used before the full
calculation was available are described in Appendix \ref{sect:nnlo}.
Note that the QCD colour factors introduced in Sect.~\ref{sec:qcd}
cannot be modified if the interpolated NNLO splitting functions
are used, since these expressions use the default QCD values.

%......................................................................
\subsubsection{Mass threshold matrices}
\label{sec:mtm}

Still in the \ttt{dglap\_objects} module, we have a type dedicated to
crossing heavy quark mass thresholds.
\begin{lstlisting}
  type(grid_def)           :: grid
  type(mass_threshold_mat) :: MTM_NNLO

  call InitMTMNNLO(grid, MTM_NNLO)   ! MTM_NNLO is coeff of (as/2pi)**2
\end{lstlisting}
This is the coefficient  of $(\as/2\pi)^2$ for the convolution matrix that
accounts for crossing a heavy flavour threshold in $\MSbar$
factorisation scheme, at $\mu_F = m_h$, where
$m_h$ is the heavy-quark pole mass, as has
been described in Sect.~\ref{sec:pqcd}. Since the
corresponding NLO term is zero, the number of flavours in $\as$ is
immaterial at NNLO. 

The treatment of $n_f$ in the \ttt{mass\_threshold\_mat} is very
specific because at NNLO, the only order in the $\MSbar$ factorisation
scheme at which it's non-zero and currently known, it is independent of $n_f$.
Its action does of course however depend on $n_f$. Since, as for
\ttt{split\_mat} objects, we don't want to the action of of the
\ttt{mass\_threshold\_mat} to depend on the availability of the
current $n_f$ information from the \ttt{qcd} module, instead we
require that before using a \ttt{mass\_threshold\_mat}, you should
explicitly indicate the number of flavours (defined as including the
new heavy flavour). This is done using a call to the
\ttt{SetNfMTM(MTM\_NNLO,nf\_incl\_heavy)} subroutine. So for example
to take a PDF in the effective theory with
$n_f=3$ active flavours \ttt{pdf\_nf3}, and convert it
to the corresponding PDF in the effective theory with
$n_f=4$ active flavours \ttt{pdf\_nf4} at $m_h^2$, one
uses code analogous to the following
\begin{lstlisting}
  real(dp) :: pdf_nf3(:,:), pdf_nf4(:,:)
  [ ... ]
  call SetNfMTM(MTM, 4)
  pdf_nf4 = pdf_nf3 + (as2pi)**2 * (MTM_NNLO.conv.pdf_nf3)
\end{lstlisting}
The convolution only works if the \ttt{pdf}'s are in the \ttt{human}
representation and an error is given if this is not the case. Any
heavy flavour (like for example intrinsic charm) 
present in \ttt{pdf\_nf3} would be left in unchanged.

Note that the type {\tt mass\_threshold\_mat}
is not currently suitable for general changes of
flavour-number. For example if you wish to carry out a change in the
DIS scheme or at a scale $\mu_F \neq m_h$ then you have
to combine a series of different convolutions (essentially correcting
with the lower number of flavours to the $\MSbar$ factorisation scheme
at $\mu_F = m_h$ before changing the number of flavours
and then correcting back to the original scheme and scale using the
higher number of flavours).

As for the NNLO splitting functions, the mass threshold corrections
come in exact and parametrised variants. By default it is the latter
that is used (provided by Vogt \cite{VogtMTMParam}).  The cost of
initialising with the exact variants of the mass thresholds is much
lower than for the exact NNLO splitting functions (partly because
there is no $n_f$ dependence, partly because it is only one flavour
component of the mass-threshold function that is complex enough to
warrant parametrisation). The variant can be chosen by the user
before initialising the \ttt{mass\_threshold\_mat} by making the
following subroutine call:
\begin{lstlisting}
  integer :: threshold_variant
  call dglap_Set_nnlo_nfthreshold(threshold_variant)
\end{lstlisting}
with the following variants defined (as integer parameters), again in
the module \ttt{dglap\_choices}:
\begin{lstlisting}
  nnlo_nfthreshold_exact
  nnlo_nfthreshold_param                 [default]
\end{lstlisting}

%......................................................................
\subsubsection{Putting it together: \ttt{dglap\_holder}}

The discussion so far in this subsection was intended to provide the
reader with an overview of the different DGLAP components that have
been implemented and of how they can be initialised individually. This
is useful above all if the user needs to tune the program to some
specific unusual application.

In practice, one foresees that most users will need just a standard
DGLAP evolution framework, and so will prefer not need to manage
all these components individually. Accordingly \hoppet provides a
type, $\ttt{dglap\_holder}$ which holds all the components 
required for
a given kind of evolution. To initialise all information for a
fixed-flavour number evolution, one does as follows
\begin{lstlisting}
  use hoppet_v1
  type(dglap_holder) :: dglap_h
  integer            :: factscheme, nloop, nf_lcl  

  nloop      = 3                 ! NNLO
  factscheme = factscheme_MSbar  ! or: factscheme_DIS; factscheme_PolMSbar
  nf_lcl = 4
  call qcd_SetNf(nf_lcl)         ! set the fixed number of flavours
  ! call qcd_SetGroup(...)       ! if you want different colour factors

  ! now do the initialisation
  call InitDglapHolder(grid, dglap_h, factscheme, nloop)
\end{lstlisting}
The constants \ttt{factscheme\_*} are defined in \ttt{module
  dglap\_choices}. %
The corrections to the splitting functions to get the DIS scheme are
implemented by carrying out appropriate convolutions of the $\MSbar$
splitting and coefficient functions. Currently the DIS scheme is only
implemented to NLO\footnote{It's NNLO implementation would actually be fairly
straightforward given the parametrisation provided in
\cite{White:2005wm}, and may be performed in
future releases of \hoppet.}.
 The polarised splitting functions are only currently known to NLO.

Initialisation can also be carried out with a single call for a range of
different numbers of flavours:
\begin{lstlisting}
  integer :: nflo, nfhi
  [...]
  nflo = 3; nfhi = 6    ! [calls to qcd_SetNf handled automatically]
  call InitDglapHolder(grid, dglap_h, factscheme, nloop, nflo, nfhi)
\end{lstlisting}
Mass thresholds are not currently correctly supported in the DIS
scheme.

For all the above calls, at NNLO the choice of exact of parametrised
splitting functions and mass thresholds is determined by the calls to
\ttt{dglap\_Set\_nnlo\_splitting} and
\ttt{dglap\_Set\_nnlo\_nfthreshold}, as described in
Sects.~\ref{sec:dglap-split} and \ref{sec:mtm} respectively. These
calls must be made prior to the call to \ttt{InitDglapHolder}.

Having initialised a \ttt{dglap\_holder} one has access to various components:
\begin{lstlisting}
  type dglap_holder
    type(split_mat), pointer :: allP(1:nloop, nflo:nfhi)  ! FFNS: nflo=nfhi=nf_lcl
    type(split_mat), pointer :: P_LO, P_NLO, P_NNLO
    type(mass_threshold_mat) :: MTM2
    logical                  :: MTM2_exists
    integer                  :: factscheme, nloop
    integer                  :: nf
    [ ... ]
  end type dglap holder
\end{lstlisting}
Some just record information passed on initialisation, for example
\ttt{factscheme} and \ttt{nloop}. Other parts are set up once and for
all on initialisation, notably the \ttt{allP} matrix, which contains
the 1-loop, 2-loop, etc. splitting matrices for the requested $n_f$
range.

Yet other parts of the \ttt{dglap\_holder} type depend on $n_f$.
Before accessing these, one should first perform the following call:
\begin{lstlisting}
  call SetNfDglapHolder(dglap_h, nf_lcl)
\end{lstlisting}
This creates links:
\begin{lstlisting}
  dglap_h%P_LO   => dglap_h%allP(1,nf_lcl)
  dglap_h%P_NLO  => dglap_h%allP(2,nf_lcl)
  dglap_h%P_NNLO => dglap_h%allP(3,nf_lcl)
\end{lstlisting}
for convenient named access to the various splitting matrices, and it
also sets the global (\ttt{qcd}) $n_f$ value (via a call to
\ttt{qcd\_SetNf}) and where relevant updates the internal $n_f$ value
associated with \ttt{MTM2} (via a call to \ttt{SetNfMTM}).

As with other types that allocate memory for derived types, that
memory can be freed via a call to the \ttt{Delete} subroutine,
\begin{lstlisting}
  call Delete(dglap_h)
\end{lstlisting}

%======================================================================
\section{DGLAP evolution}
So far we have described all the tools that are required
to perform  DGLAP convolutions of PDFs. In this section
we describe how the different ingredients are put together
to perform the actual evolution.

%----------------------------------------------------------------------
\subsection{Running coupling}
\label{sec:run-coupl}

Before carrying out any DGLAP evolutions, one first needs to set up 
a \ttt{running\_coupling} object (defined in \ttt{module
  qcd\_coupling}):
\begin{lstlisting}
  type(running_coupling) :: coupling
  real(dp) :: alfas, Q, quark_masses(4:6), muMatch_mQuark
  integer  :: nloop, fixnf

  [... set parameters ...]
  call InitRunningCoupling(coupling [, alfas] [, Q] [, nloop] [, fixnf]&
                          & [, quark_masses] [, muMatch_mQuark])
\end{lstlisting}
As can be seen, many of the arguments are optional. Their default
values are as follows:
\begin{lstlisting}
  Q     = 91.2_dp
  alfas = 0.118_dp  ! Value of coupling at scale Q
 
  nloop = 2
  fixnf = [.not. present]
                       !     charm,      bottom,   top
  quark_masses(4:6) = (/ 1.414213563_dp, 4.5_dp, 175.0_dp /) ! Heavy quark pole masses
  muMatch_mQuark    = 1.0_dp
\end{lstlisting}
The running coupling object is initialised so that at scale \ttt{Q}
the coupling is equal to \ttt{alfas}. The running is carried out with
the \ttt{nloop} $\beta$-function. If the \ttt{fixnf} argument is
present, then the number of flavours is kept fixed at that
value. Otherwise flavour thresholds are implemented at scales
\begin{lstlisting}
  muMatch_mQuark * quark_masses(4:6)
\end{lstlisting}
where the quark masses are \emph{pole} masses. The choice to use pole
masses (and their particular default values) is inspired by the
benchmark runs~\cite{Benchmarks} in which \hoppet results were
compared to those of Vogt's moment-space code
QCD-Pegasus~\cite{Pegasus}. The default value of the
QCD coupling is taken to be close
to the present world average \cite{Bethke:2006ac}.

To access the coupling at some scale \ttt{Q} one uses the following
function call:
\begin{lstlisting}
  alfas = Value(coupling, Q [, fixnf])
\end{lstlisting}
This is the value of the coupling as obtained from the Runge-Kutta
solution of the \ttt{nloop} version of eq.~(\ref{eq:as-ev}) (the
numerical solution is actually carried out for $1/\as$), together with the
appropriate mass thresholds. For
typical values of $\as(M_Z)$ the coupling is guaranteed to be reliably
determined in the range $0.5 \GeV < Q < 10^{19}\GeV$. The values of
the $\beta$ function coefficients used in the evolution correspond to
those obtained with the values of the QCD colour factors that were in
vigour at the moment of initialisation of the coupling.

In the variable flavour-number case, the \ttt{fixnf} argument allows
one to obtain the coupling for \ttt{fixnf} flavours even outside the
natural range of scales for that number of flavours. This is only
really intended to be used close to the natural range of scales, and
can be slow if one goes far from that range (a warning message will be
output). If one is interested in a coupling that (say) never has more
than 5 active flavours, then rather than using the \ttt{fixnf} option
in the \ttt{Value} subroutine, it is best to initialise the coupling
with a fictitious large value for the top mass.

Often it is convenient to be able to enquire about the mass information
embodied in a \ttt{running\_coupling}. For example in the PDF
evolution below, all information about the location of mass thresholds is
obtained from the \ttt{running\_coupling} type.

The quark pole mass for flavour \ttt{iflv} can be obtained with the
call
\begin{lstlisting}
  pole_mass = QuarkMass(coupling, iflv)
\end{lstlisting}
The range of scales, $\ttt{Qlo} < Q < \ttt{Qhi}$ for which \ttt{iflv}
is the heaviest active flavour is obtained by the subroutine call
\begin{lstlisting}
  call QRangeAtNf(coupling, iflv, Qlo, Qhi [, muM_mQ])
\end{lstlisting}
The optional argument \ttt{muM\_mQ} allows one to obtain the answer as
if one had initialised the coupling with a different value of
\ttt{muMatch\_mQuark} than that actually used. One can also establish
the number of active flavours, \ttt{nf\_active}, at a given scale
\ttt{Q} with the following function:
\begin{lstlisting}
  nf_active = NfAtQ(coupling, Q [, Qlo, Qhi] [, muM_mQ])
\end{lstlisting}
As well as returning the number of active flavours, it can also set
\ttt{Qlo} and \ttt{Qhi}, which correspond to the range of scales in
which the number of active flavours is unchanged. The optional
\ttt{muM\_mQ} argument has the same purpose as in the \ttt{QRangeAtNf}
subroutine. The last of the enquiry functions allows one to obtain the
range of number of flavours covered in this coupling, $\ttt{nflo} \le
n_f \le \ttt{nfhi}$:
\begin{lstlisting}
  call NfRange(coupling, nflo, nfhi)
\end{lstlisting}

Finally, as usual, once you no longer need a \ttt{running\_coupling}
object, you may free the memory associated with it using the
\ttt{Delete} call:
\begin{lstlisting}
  call Delete(coupling)
\end{lstlisting}

%----------------------------------------------------------------------
\subsection{DGLAP evolution}
\label{sec:dglap-ev}

%......................................................................
\subsubsection{Direct evolution}
\label{sec:direct-evolution}

We are now, at last,
 ready to evolve a multi-flavour PDF. This is done by breaking the
evolution into steps, and for each one using a Runge-Kutta
approximation for the solution of a first-order matrix differential
equation. The steps are of uniform size in a variable $u$ that
satisfies the following approximate relation
\begin{equation}
  \label{eq:du}
  \frac{du}{d\ln Q^2} \simeq \as(Q^2) \ .
\end{equation}
For a 1-loop running coupling one has $u = (\ln \ln
Q^2/\Lambda)/\beta_0$, which is the variable that appears in
analytical solutions to the $1$-loop DGLAP equation.
The step size in u, du, can be set with the following call
\begin{lstlisting}
  real(dp) :: du = 0.1_dp ! or some smaller value
  call SetDefaultEvolutionDu(du)
\end{lstlisting}
The error on the evolution from the finite step size should scale as
$(\ttt{du})^4$. With the default value of $\ttt{du}=0.1$, errors are
typically somewhat smaller than $10^{-3}$ (see
Sect.~\ref{sec:benchmarks} for the detailed benchmarks).

To actually carry out the evolution, one uses the following subroutine
call:
\begin{lstlisting}
  type(dglap_holder)     :: dglap_h
  type(running_coupling) :: coupling
  real(dp), pointer      :: initial_pdf(:,:)
  real(dp)               :: Q_init, Q_end
  integer                :: nloop
  integer                :: untie_nf
  [...]
  call EvolvePDF(dglap_h, initial_pdf, coupling, Q_init, Q_end &
                        & [, muR_Q] [, nloop] [, untie_nf] [, du] )
\end{lstlisting}
which takes a PDF array 
\ttt{pdf} and uses the splitting matrices in \ttt{dglap\_h}  to
evolve it from scale \ttt{Q\_init} to scale \ttt{Q\_end}.
By default the renormalisation to factorisation scale ratio is
$\ttt{muR\_Q} = 1.0$ and the number of loops in the evolution is the
same as was used for the running \ttt{coupling} (the \ttt{nloop}
optional argument makes it possible to override this choice). Variable
flavour-number switching takes place at the pole masses (maybe one day
this restriction will be lifted) as associated with the
\ttt{coupling}. 

If the \ttt{dglap\_holder} object \ttt{dglap\_h} does not support the
relevant number of loops or flavours, the program will give an error
message and stop. With the \ttt{untie\_nf} option you can request that
the number of flavours in the evolution be `untied' from that in the
coupling in the regions where \ttt{dglap\_h} does not support the number of
flavours used in the coupling. Instead the closest number of flavours
will be used.\footnote{For example if \ttt{dglap\_h} was initialised with
  $n_f = 3\ldots5$ while the coupling has $n_f = 3\ldots 6$, then
  variable flavour number evolution will be used up to $n_f = 5$, but
  beyond the top mass the evolution will carry on with $5$ flavours,
  while the coupling uses $6$ flavours. There probably aren't too many
  good reasons for doing this (other than for examining how much it
  differs from a `proper' procedure).}

Mass thresholds (NNLO) are implemented as described in
Sect.~\ref{sec:pqcd}:
\begin{subequations}
\label{eq:mass_threshold}
\begin{align}
  \ttt{pdf}_{n_f} &= \ttt{pdf}_{n_f-1} +
  \left(\frac{\alpha_{s}^{(n_f)}(x_\mu m_h^2)}{2\pi}\right)^2
  (\mbox{\ttt{dglap\_h\%MTM2 .conv.}}\; \ttt{pdf}_{n_f-1}) \ ,\\
  \ttt{pdf}_{n_f-1} &= \ttt{pdf}_{n_f} \;\;\;\;-
  \left(\frac{\alpha_{s}^{(n_f)}(x_\mu m_h^2)}{2\pi}\right)^2
  (\mbox{\ttt{dglap\_h\%MTM2 .conv.}}\; \ttt{pdf}_{n_f}) \ ,
\end{align}
\end{subequations}
when crossing the threshold upwards and downwards, respectively. Note
that the two operations are not perfect inverses of each other,
because the number of flavours of the \ttt{pdf} used in the
convolution differs in the two cases. The mismatch however is only of
order $\as^4$ (NNNNLO), \ie well beyond currently known accuracies.

A general remark is that crossing a flavour threshold downwards will
nearly always result in some (almost certainly physically spurious)
intrinsic heavy-flavour being left over below threshold.

%......................................................................
\subsubsection{Precomputed evolution and the \ttt{evln\_operator}}
\label{sec:precomputed-evolution}

Each Runge-Kutta evolution step involves multiple evaluations of the
derivative of the PDFs, and the evolution between two scales may be
broken up into multiple Runge-Kutta steps. This amounts to a large
number of convolutions. It can therefore be useful to create a single
\emph{derived} splitting matrix that is equivalent to the whole
evolution between the two scales. 

A complication arises because evolutions often cross flavour
thresholds, whereas a derived splitting matrix is only valid for fixed
$n_f$. Therefore a new type has to be created, \ttt{evln\_operator},
which consists of a linked list of splitting and mass threshold
matrices, breaking an evolution into a chain of interleaved
fixed-flavour evolution steps and flavour changing steps. An
\ttt{evln\_operator} is created with a call that is almost identical
to that used to evolve a PDF:
\begin{lstlisting}
  type(evln_operator) :: evop
  real(dp), pointer   :: pdf_init(:,:), pdf_end(:,:)
  [...]
  call InitEvlnOperator(dglap_h, evop, coupling, Q_init, Q_end &
                           & [, muR_Q] [, nloop] [, untie_nf] [, du] )
\end{lstlisting}
It can then be applied to PDF in the same way that a normal
\ttt{split\_mat} would:
\begin{lstlisting}
  pdf_end = evop * pdf_init       ! assume both pdfs already allocated
  ! OR (alternative form) 
  pdf_end = evop .conv. pdf_init  
\end{lstlisting}
As usual the \ttt{Delete} subroutine can be used to clean up any
memory associated with an evolution operator that is no longer needed.

%======================================================================
%======================================================================
\section{Tabulated PDFs}
\label{sec:tabulated-pdfs}

The tools in the previous section are useful if one knows that one
needs DGLAP evolution results at a small number of predetermined $Q$
values. Often however one simply wishes to provide a PDF distribution
at some initial scale and then subsequently be able to access it at
arbitrary values of $x$ and $Q$. For this purpose it is useful 
(and most efficient) to
produce a table of the PDF as a function of $Q^2$, which then allows
for access to the PDF at arbitrary $x, Q$ using an interpolation. All
types and routines discussed in this section are in the
\ttt{pdf\_tabulate} module, or accessible also from \ttt{hoppet\_v1}.

\subsection{Preparing a PDF table}
The type that contains a PDF table is \ttt{pdf\_table}. It first needs
to be allocated,
\begin{lstlisting}
  type(pdf_table) :: table
  [...]

  call AllocPdfTable(grid, table, Qmin, Qmax &
                     & [, dlnlnQ ] [, lnlnQ_order ] [, freeze_at_Qmin] )
\end{lstlisting}
where one specifies the range of Q values to be tabulated, from
\ttt{Qmin} to \ttt{Qmax}, and optionally the interpolation step size
\ttt{dlnlnQ} in the variable $\ln \ln Q/(0.1\GeV)$ (default
$\ttt{dlnlnQ}=0.07$, sufficient for $10^{-3}$ accuracy), the
interpolation order \ttt{lnlnQ\_order}, equal to $3$ by default, and
finally whether PDFs are to be frozen below \ttt{Qmin}, or instead set
to zero (the default is \ttt{freeze\_at\_Qmin=.false.}, \ie they are
set to zero).\footnote{Note that the spacing and interpolation in $Q$
  are treated independently of what's done in the PDF evolution. A
  reason for this is that in the PDF evolution one uses a variable
  related to the running coupling, which is similar to $\ln \ln Q$ but
  whose precise details depend on the particular value of the
  coupling.  Using this in the tabulation would have prevented one
  from having a tabulation disconnected from any coupling.
  Unfortunately \ttt{du} and \ttt{dlnlnQ} are not normalised
  equivalently --- roughly speaking for $n_f = 4$ they correspond to
  the same spacing if $\ttt{dlnlnQ} \simeq 0.7 \ttt{du}$.}

By default a tabulation knows nothing about $n_f$ thresholds, which
means that in the neighbourhood of thresholds the tabulation would be
attempting to interpolate a discontinuous function (at NNLO). To
attribute information about $n_f$ thresholds to the tabulation, set
them up first in a running \ttt{coupling} object and then transfer
them:
\begin{lstlisting}
  call AddNfInfoToPdfTable(table,coupling)
\end{lstlisting}
When interpolating the table (see below), the set of $Q$ values for
the interpolation will be chosen to as to always have a common $n_f$
value. Note that \ttt{AddNfInfoToPdfTable} may only be called once for
an allocated table: if you need to change the information about $n_f$
thresholds, \ttt{Delete} the table, reallocate it and then reset the
$n_f$ information. This is not necessary if you just
change the value of the coupling.

Given an existing table, \ttt{ref\_table}, a new table,
\ttt{new\_table}, can be allocated with identical properties
(including any $n_f$ information) as follows
\begin{lstlisting}
  call AllocPdfTable(new_table, ref_table)
\end{lstlisting}
All of the above routines can be used with $1$-dimensional arrays of
tables as well (in \ttt{AllocPdfTable} the reference table must always
be a scalar).

A table can be filled either from a routine that provides the PDFs as
a function of $x$ and $Q$, or alternatively by evolving a PDF at an
initial scale. The former can be achieved with
\begin{lstlisting}
  call FillPdfTable_LHAPDF(table, LHAsub)
\end{lstlisting}
where \ttt{LHAsub} is any subroutine with the LHAPDF interface,
{\it i.e.}, as shown earlier
in Sect.~\ref{sec:human-rep}.

To fill a table via an evolution from an initial scale, one uses 
\begin{lstlisting}
  type(pdf_table)        :: table
  type(dglap_holder)     :: dglap_h
  type(running_coupling) :: coupling
  real(dp), pointer      :: initial_pdf(:,:)
  real(dp)               :: Q0
  integer                :: nloop
  integer                :: untie_nf
  [...]
  call EvolvePdfTable(table, Q0, initial_pdf, dglap_h, coupling &
                      & [, muR_Q] [, nloop] [, untie_nf] )
\end{lstlisting}
which takes an the \ttt{initial\_pdf} at scale \ttt{Q0}, and evolves
it across the whole range of $Q$ values in the table, using the
\ttt{EvolvePDF} routine. The arguments have the same meaning as
corresponding ones in \ttt{EvolvePDF}, explained in
Sect.~\ref{sec:direct-evolution}. The $\ttt{du}$ value that's used
is the default one for \ttt{EvolvePDF} which, we recall, may be set
using \ttt{SetDefaultEvolutionDu(du)}. If the $Q$ spacing in the
tabulation is such that steps in \ttt{du} would be too large, then the
steps are automatically resized to the tabulation spacing.

One may also use the precomputed evolution facilities of
Sect.~\ref{sec:precomputed-evolution}, by calling the routine
\begin{lstlisting}
  call PreEvolvePdfTable(table, Q0, dglap_h, coupling, &
                         & [, muR_Q] [, nloop] [, untie_nf] )
\end{lstlisting}
prepares \ttt{evln\_operator}s for all successive $Q$ intervals in the
table. An accelerated evolution that uses these operators instead of
explicit Runge-Kutta steps may then be obtained by calling
\begin{lstlisting}
  call EvolvePdfTable(table, initial_pdf)
\end{lstlisting}
The \ttt{EvolvePdfTable} routine may be called as many times as one
likes, for different initial PDFs for example;
however, if one wishes to change the parameters of
the evolution (coupling, perturbative order, etc.) 
in the precomputed option, one must
first \ttt{Delete} the table and then prepare again it.

%......................................................................
\subsection{Accessing a table}
\label{sec:acc_table}

The main way to access a table is as follows
\begin{lstlisting}
real(dp) :: pdf(-6:6), y, x, Q
  [...]
  call EvalPdfTable_yQ(table, y, Q, pdf)
  ! or using x
  call EvalPdfTable_xQ(table, x, Q, pdf)
\end{lstlisting}
There may be situations where it is useful to access the internals of
a \ttt{pdf\_table}, for example because one would like to carry out a
convolution systematically on the whole contents of the table. Among
the main elements are
\begin{lstlisting}
  type pdf_table
    integer           :: nQ           ! arrays run from 0:nQ
    real(dp), pointer :: tab(:,:,:)   ! the actual tabulation
    real(dp), pointer :: Q_vals(:)    ! the Q values 
    integer,  pointer :: nf_int(:)    ! nf values at each Q
    real(dp), pointer :: as2pi(:)     ! alphas(Q)/2pi at each Q
    [...]
  end type pdf_table
\end{lstlisting}
where the third dimension of \ttt{tab} spans the range of tabulated
$Q$ values, and the $n_f$ and coupling information are only allocated
and set if one has called \ttt{AddNfInfoToPdfTable} for the table.

An example of usage of the low-level information contained in the
table is the following, which initialises \ttt{table\_deriv\_LO} with
the LO derivative of \ttt{table}:
\begin{lstlisting}
  do iQ = 0, table%nQ
    table_deriv_LO%tab(:,:,iQ) = table%as2pi(iQ) * &
                     & ( dglap_h%allP(1,table%nf_int(iQ)) * table%tab(:,:,iQ))
  end do
\end{lstlisting}
where we assume \ttt{table\_deriv\_LO} to have been allocated with an
appropriate structure at some point, \eg via 
\begin{lstlisting}
  call AllocPdfTable(table_deriv_LO, table)
\end{lstlisting}
The above mechanism has found use in the a-posteriori PDF
library~\cite{APPL,Banfi:2007gu} 
and in work matching event shapes with fixed-order
calculations~\cite{caesar,DisResum}.
One could also imagine using it to obtain tables of
(flavour-separated) structure functions, if one were to convolute with
coefficient functions rather than splitting functions.

As with all other objects, a \ttt{pdf\_table} object can be deleted
using
\begin{lstlisting}
  call Delete(table) 
\end{lstlisting}
Notice that a table is a local variable in each procedure and so 
effectively a different variable each separate
procedure. 
If ones needs to use a 
PDF table across different procedures, it has to be 
defined within a module.
In Appendix \ref{sec:example_program}, 
there is a detailed example of different ways of
accessing a table.

%======================================================================
%======================================================================
\section{Streamlined interface}
\label{sec:vanilla}

Now we present the streamlined interface to \hoppet,
intended to allow easy access
to the essential evolution functionality from languages other than
F95.  It hides all the object-oriented nature of the program, and
provides access to one \ttt{pdf\_table}, based on a single grid
definition. The description will be given as if one is calling from
F77. An include file \ttt{src/hoppet\_v1.h} is provided for calling
from {\tt C++} --- the interface is essentially identical to the Fortran
one, with the caveat that names are case sensitive (the cases are as
given below), and that PDF components referred to below as
\ttt{pdf(-6:6)} become \ttt{pdf[0..12]}. A summary of the
most relevant procedures of this interface 
and their description can be found 
in the reference guide, Appendix~\ref{sec:refguide}.

%......................................................................
\subsection{Initialisation}
\label{sec:vanilla_initialisation}

The simplest way of initialising the 
streamlined interface is by calling
\begin{lstlisting}
  call hoppetStart(dy,nloop)
\end{lstlisting}
which will set up a compound (four different spacings) grid with
spacing \ttt{dy} at small $x$, extending to $y = 12$, and numerical
order $\ttt=-5$.  The $Q$ range for the tabulation will be $1\GeV <
Q<28 \TeV$ and a reasonable choice will be made for the \ttt{dlnlnQ}
spacing (related to \ttt{dy}). It will initialise splitting functions
up to \ttt{nloop} loops (though one can still carry out evolutions
with fewer loops).
If you need more control over the initialisation, you should use
\begin{lstlisting}
  call hoppetStartExtended(ymax,dy,Qmin,Qmax,dlnlnQ,nloop,order,factscheme)
\end{lstlisting}
which will again set up compound grid, but give control over the
numerical \ttt{order} and the $y$ and $Q$ ranges and spacings (as
before \ttt{dy} is the spacing at small $x$). It also allows one to
choose the type of evolution according to \ttt{factscheme}.

%......................................................................
\subsection{Usage}
\label{sec:vanilla_usage}

To carry out an evolution, one should first decide whether one wants a
fixed-flavour number scheme or a variable flavour number scheme (the
default). Either can be set with its parameters as follows:
\begin{lstlisting}
  call hoppetSetFFN(fixed_nf)
  call hoppetSetVFN(mc, mb, mt) ! Heavy quark pole masses
\end{lstlisting}
where for the VFN one specifies the pole-masses for the quarks. An
evolution is carried out with the following routine
\begin{lstlisting}
  call hoppetEvolve(asQ, Q0alphas, nloop, muR_Q, LHAsub, Q0pdf)
\end{lstlisting}
where one specifies the coupling \ttt{asQ} at a scale \ttt{Q0alphas},
the number of loops for the evolution, \ttt{nloop}, the ratio of
renormalisation to factorisation scales \ttt{muR\_Q},\footnote{Note
  that in the streamlined interface, with $\ttt{muR\_Q}\ne 1$, the running
  coupling flavour thresholds are still placed at the quark masses;
  the evolution needs the coupling for a given number of flavours
  outside the standard range for that number of flavours (precisely
  because $\ttt{muR\_Q}\ne 1$) and this is done automatically in the
  evolution. In contrast in the benchmark studies~\cite{Benchmarks},
  the flavour thresholds for the coupling were placed at $\ttt{muR\_Q}
  \times m_Q$.  This is a perfectly valid alternative, but can
  complicate the specification of the $\as$ value --- for example with
  $\ttt{muR\_Q} = 0.5$ the matching for the top threshold would be
  carried out at $\mu = 0.5 m_t \simeq 85 \GeV$, and if one specified
  the coupling at scale $M_Z$, it wouldn't be clearer whether this was
  a $5$-flavour value or a $6$-flavour value. With the procedure
  adopted in the streamlined interface the issue does not arise. (While in
  F95 the user has the freedom to do as they prefer). } %
the name of a subroutine \ttt{LHAsub} with interface
\begin{lstlisting}
  subroutine LHAsub(x,Q,pdf)
    implicit none
    double precision x,Q,pdf(-6:6)
    [...]   ! sets pdf to be momentum densities, e.g. pdf(0) = xg(x)
  end subroutine 
\end{lstlisting}
to return the initial condition for the evolution and the scale
\ttt{Q0pdf} at which one starts the PDF evolution. Note that the
LHAsub subroutine will only be called with $\ttt{Q}=\ttt{Q0pdf}$.
To access the coupling one uses
\begin{lstlisting}
  alphas = hoppetAlphaS(Q)
\end{lstlisting}
while the PDF at a given value of $\ttt{x}$ and $\ttt{Q}$ is obtained with 
\begin{lstlisting}
  call hoppetEval(x,Q,f)
\end{lstlisting}
which sets \ttt{f(-6:6)} (recall that it is $xg(x)$, etc., that are
returned, since this is what is
used through \hoppet).

It is also possible to prepare an evolution in cached form. This is
useful if one needs to evolve many different PDF sets with the same
evolution properties (coupling, initial scale, etc.), as is the
usual situation in global analyses of PDFs, because though
the preparation may take a bit longer than a normal evolution
($2$--$10$ times depending on the \ttt{order}), once it is done,
cached evolutions run $3$--$4$ faster than a normal evolution. The
preparation of the cache is carried out with
\begin{lstlisting}
  call hoppetPreEvolve(asQ, Q0alphas, nloop, muR_Q, Q0pdf)
\end{lstlisting}
and then the cached evolution is carried out with 
\begin{lstlisting}
  call hoppetCachedEvolve(LHAsub)
\end{lstlisting}
The results may be very slightly different from those in a normal
evolution (some information is lost when caching), and the user 
may wish to check on a case-by-case basis that such differences
don't matter in practice.

The tabulation can also be filled with the contents of an external PDF
package, by calling
\begin{lstlisting}
  call hoppetAssign(LHAsub)
\end{lstlisting}
where \ttt{LHAsub} is the name of any subroutine with the interface
given above, which will now be called with a range of \ttt{Q} values
corresponding to the internal tabulation scales. This essentially just
transfers an external tabulation into \hoppet's internal
representation. 

Finally given an evolved or assigned PDF, one can obtain information
about convolutions of the splitting functions with the PDFs:
\begin{lstlisting}
  call hoppetEvalSplit(x,Q,iloop,nf,f)
\end{lstlisting}
sets \ttt{f(-6:6)} equal to the value at \ttt{x}, \ttt{Q} of
convolution of the \ttt{iloop} splitting function matrix (with
\ttt{nf} flavours) with the currently tabulated PDF. If $\ttt{nf}<0$
the number of flavours used is the one appropriate at the specified
$Q$ scale (as long as the information is available, \ie one of
\ttt{hoppetEvolve} or \ttt{hoppetCachedEvolve} has been called). The
first call with a given \ttt{nf} for a specified \ttt{iloop} will be
slow ($\sim$ the time for a cached evolution), but subsequent calls
with the same values will be fast.

The routines described here 
%(and associated data storage) 
are to be 
found in \ttt{src/streamlined\_interface.f90} and may provide inspiration
for the user wishing to write their own F95 code for \hoppet.

%======================================================================
%======================================================================
\section{Benchmarks}
\label{sec:benchmarks}

Key questions in assessing the usefulness of a PDF evolution code include
that of its correctness, its accuracy and its speed. \hoppet's
correctness has been established with a reasonable degree of
confidence in the benchmark tests~\cite{Benchmarks} where it was
compared with the Mellin space based
evolution code QCD-Pegasus \cite{Pegasus}. 
The program used to carry out those tests
is available as \ttt{benchmarks/benchmarks.f90}. The user should
carefully read the detailed
comments at the beginning for usage instructions.

The results used in \cite{Benchmarks} were obtained with very finely
spaced grids, in order to guarantee small numerical errors ($\lesssim
10^{-7}$).  Such accuracies are useful when comparing and testing two
independent codes, because differences or bugs in the
implementation of the physics (especially the higher-order parts) may
only manifest themselves as small changes in the results.

In contrast, for use in most physical applications, an accuracy in the
range $10^{-3}$ to $10^{-4}$ is generally y more than adequate, since it
is rare for other sources of numerical uncertainty (\eg Monte Carlo
integration errors in NLO codes, or experimental errors) to be
comparably small. The critical issue in such cases is more likely to
be the speed of the code, for example in PDF fitting
applications.

\hoppet's accuracy and speed both depend on the choice of grid (in
$y$) and the evolution and/or tabulation steps in $Q$. We shall start
with the question of the accuracy.

%......................................................................
\subsection{Accuracy}
\label{sec:Accuracy}
To measure the accuracy, 
we use the same initial condition and evolution parameters as
in~\cite{Benchmarks}:
\begin{subequations}
  \label{eq:init}
  \begin{align}
    x u_v(x)   &= 5.107200 x^{0.8} (1-x)^3\,,\\
    x d_v(x)   &= 3.064320 x^{0.8} (1-x)^4\,,\\
    x\bar d(x) &= 0.1939875 x^{-0.1} (1-x)^6\,,\\
    x\bar u(x) &= x\bar d(x) (1-x)\,,\\
    x     s(x) &= x\bar s(x) = 0.2(x\bar d(x) + x\bar u(x))\,,\\
    x g(x) &= 1.7 x^{-0.1} (1-x)^5\,,
  \end{align}
\end{subequations}
where $u_v \equiv u - \bar u$, $d_v \equiv d - \bar d$, and all other
flavours are zero. The initial scale~
\footnote{With $\bar{\epsilon}$ an infinitesimal number. Note that
this is unrelated to the evolution accuracy $\epsilon$, introduced later
in this section.} is $Q_0 = \lp \sqrt{2}-\bar{\epsilon}
\rp \GeV$,
$\as(Q_0) = 0.35$ and the charm, bottom and top pole masses are kept
at the default values, $\sqrt{2}$, $4.5$ and $175\GeV$ respectively
(as used also in the streamlined interface).
Observe that the initial conditions 
and coupling are actually both given for three active
flavours (\ie infinitesimally below the charm mass). The evolution is
carried out to NNLO accuracy in a variable flavour number scheme,
including the mass thresholds in the coupling and PDF.

\begin{figure}
  \centering
  \includegraphics[width=0.7\textwidth]{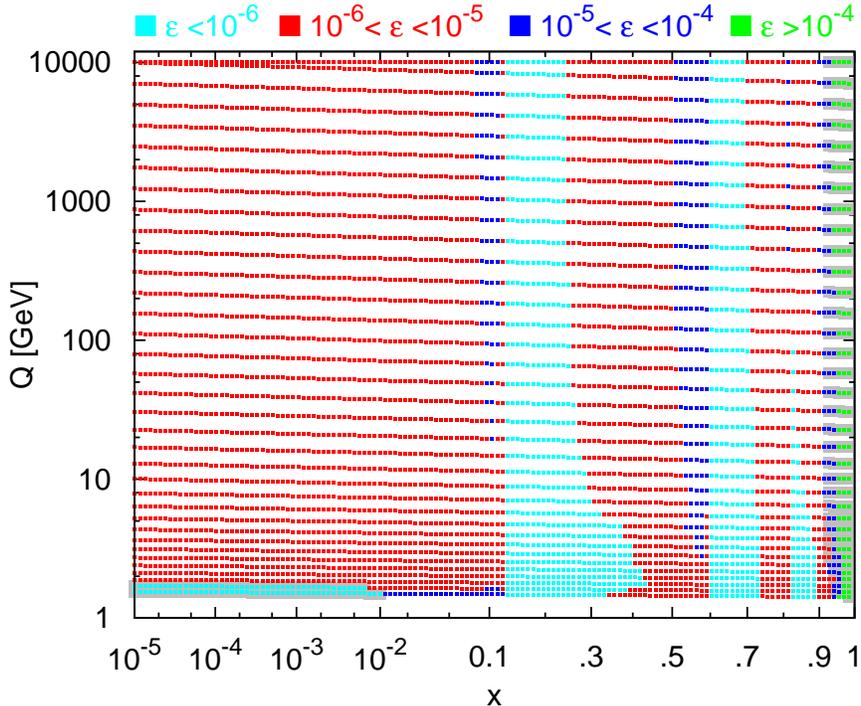}%
  \caption{The set of points in $x, Q$ used to determine the accuracy
    of the evolution. The areas shaded in grey are regions where one
    of the flavours is in the neighbourhood of a sign-change
    (bottom-left: $c,\bar c$, right: $\bar u$) and so is ignored in
    the accuracy determination.  The accuracies shown here correspond
    to a $y$ grid with a base spacing of $\ttt{dy}=0.2$ and other
    parameters as described in the text for
    Fig.~\ref{fig:acc-fixed-dy-dlnlnQ}.
    %, and
    %subsiduary grids with $3$, $9$ and $27$ times the resolution,
    %extending to $y=2.0, 0.5, 0.2$ and interpolation order $-6$; the
    %scale grid has a spacing $\ttt{dlnlnQ}=0.05$ and an interpolation
    %order of $4$ \comment{check}. 
    %
    The colour coding indicates the error in the least-well determined
    (non-excluded) flavour channel ($\bar b \ldots b$) at each point.
  }
  \label{fig:grid}
\end{figure}

All tests here are carried out based on tabulations of the PDF
evolution, Sect.~\ref{sec:tabulated-pdfs}.
We first run \hoppet with a very fine grid spacing to provide a
reference result. Then we run the evolution for a coarser grid ---
the accuracy of the coarser grid is determined by comparing its
results with those from the reference grid. We determine the relative
accuracy for each flavour at $5000$ points in the $x, Q$ plane, as
shown in Fig.~\ref{fig:grid}. The points are uniformly spaced in
$\zeta = \ln 1/x + 9(1-x)$ so as to obtain fine coverage at small and
large $x$. The $Q$ values are chosen more closely spaced at low $Q$
where the evolution is fastest and they are taken slightly correlated
with $\zeta$ so as to cover a nearly continuous range of
$Q$.\footnote{This procedure differs from that in \cite{Benchmarks}
  where fewer (500) points were used and one compared not individual
  flavours, but combinations intended to be more directly revealing of
  any deficiencies in the evolution. This reflects the difference in
  needs between obtaining a global measure of the accuracy and
  providing benchmarks intended in part to facilitate the debugging of
  independent codes.}

One difficulty that arises when examining relative accuracies is that
some flavours change sign as one varies $x$ or $Q$. Close to the zero
the relative accuracy diverges because of the small value of the
denominator. Therefore in global accuracy estimates, we eliminate
flavours in the region where they change sign (within $\Delta
\zeta=0.4$ and at the neighbouring $Q$ value). Specifically, for our
initial conditions, this corresponds to $c,\bar c$ for the two lowest
$Q$ values below $x\sim 10^{-2}$ and $\bar u$ for $x\gtrsim
0.9$.\footnote{The change in sign of the charm distribution is not
  worrying physically since it is close to threshold where it will be
  compensated for by finite mass effects in the coefficient functions;
  for $\bar u$ the sign change is more surprising, though it may be
  related to non-trivial interactions between the evolutions of the
  $u$ and $\bar u$ components at NNLO (note that in the region of the
  sign change they differ by many orders of magnitude).  } %
The exact regions are shaded in grey in Fig.~\ref{fig:grid}.

Our tabulation covers the range $10^{-5}<x<1$, $\sqrt{2} < Q<
10^4\GeV$. The grid in $y=\ln1/x$ will consist of 4 nested subgrids:
one covering the whole $y$ range with spacing $\ttt{dy}$ and others
with spacings $\ttt{dy}/3, \ttt{dy}/9, \ttt{dy}/27$ extending to
$y=2,0.5,0.2$ respectively. Except where stated we shall use
$\ttt{order}=-6$. In $Q$ the default interpolation order will be $4$.
The reference grids use $\dy=0.025$ and $\dlnlnQ=0.005$.

\begin{figure}
  \centering
  \includegraphics[width=0.48\textwidth]{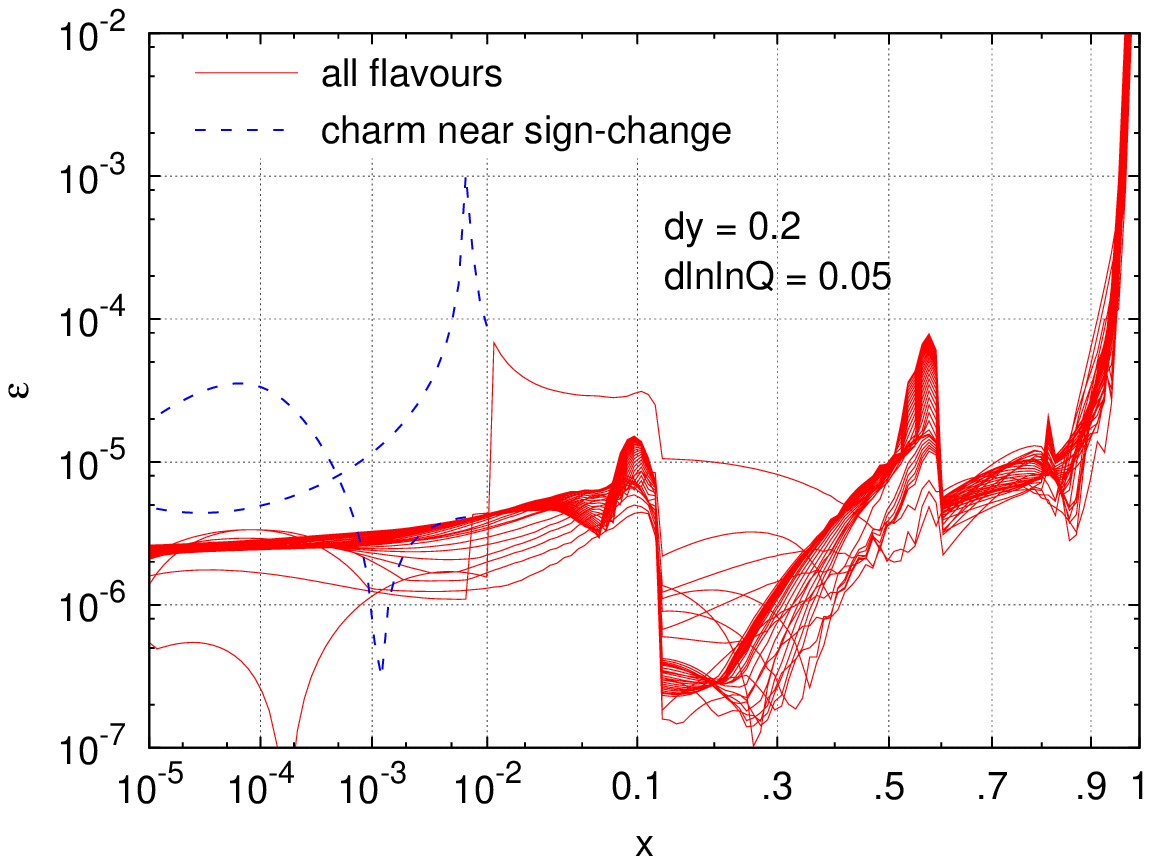}\hfill
  \includegraphics[width=0.48\textwidth]{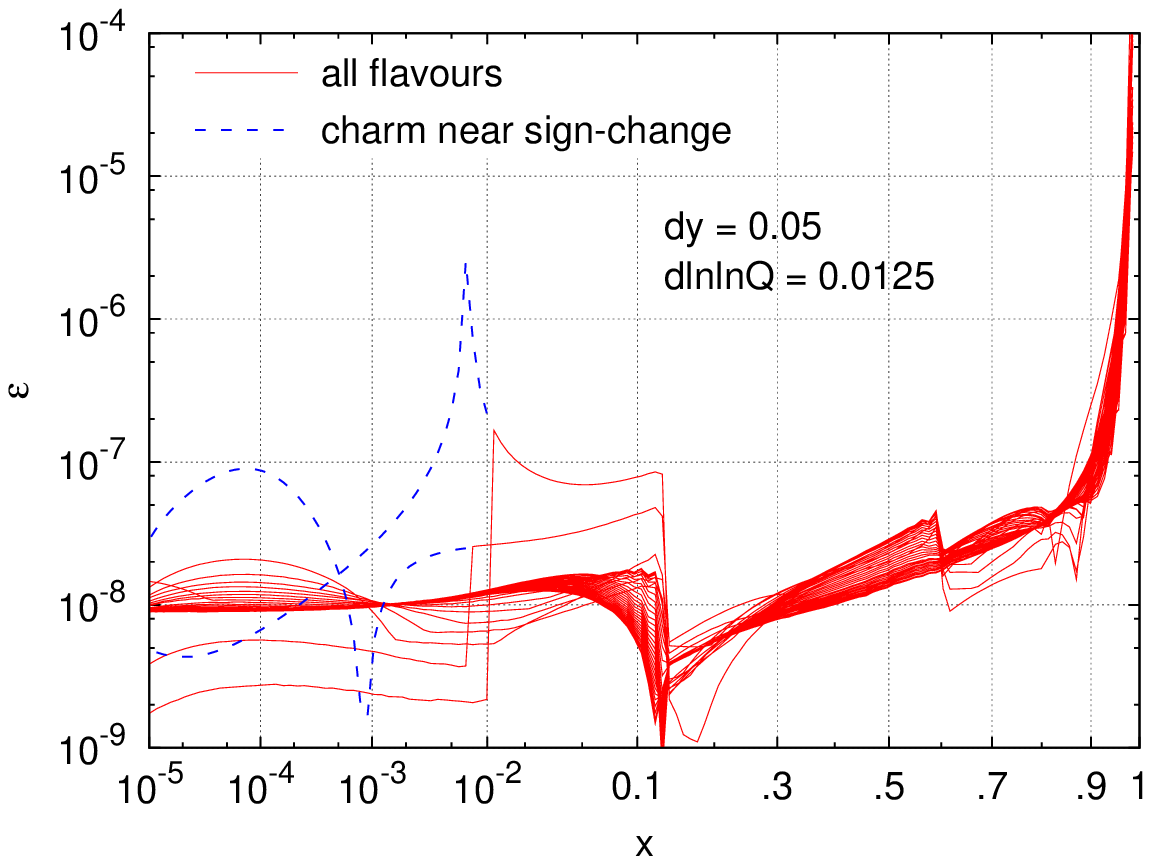}
  \caption{The relative accuracy $\epsilon$ of the least well
    determined flavour channel at each $x, Q$ point, shown as a
    function of $x$ for many $Q$ values. The results for the part of
    the charm distribution excluded from the analysis (near sign
    change) are shown separately. }
  \label{fig:acc-fixed-dy-dlnlnQ}
\end{figure}

Fig.~\ref{fig:acc-fixed-dy-dlnlnQ} shows the relative accuracy
$\epsilon$ as a function of $x$ for two grid-spacing choices (left
$\dy=0.2$, right $\dy=0.05$, $\dlnlnQ=\dy/4$ in both cases). Each
solid line corresponds to one $Q$ value and shows the error in the
least-well-determined flavour channel at each $x$, excluding flavour
channels close to a sign-change. The relative accuracy $\epsilon$
is poorest as
one approaches $x=1$, where the PDFs all go to zero very rapidly and
so have divergent logarithmic derivatives in $x$, $d\ln q/d\ln x$,
adversely affecting the accuracy of the convolutions. This region is
always the most difficult in $x$-space methods, however the use of
multiple subgrids in $x$ allows to one to obtain acceptable results
for $x<0.9$ which is likely to be the largest value of any
phenomenological relevance. 

At $x\sim0.1$, $0.6$ and $0.8$ one notices
step-like structures --- these are the points where one switches
between subgrids, with a significant degradation in accuracy at $x$
values below the transition. These structures are also visible in the
colour-coded accuracy representation in Fig.~\ref{fig:grid}, which
corresponds to $\dy=0.2$ and allows one to visualise more clearly the
$Q$ dependence of the accuracy. The effect of the grid spacing is
clearly visible as one goes from the left to the right-hand plots of
Fig.~\ref{fig:acc-fixed-dy-dlnlnQ},
with the reduction in the spacings by a factor of $4$ leading to an
improvement in accuracy by a factor $\sim 100$.

For completeness we also show the parts of the charm channel that have
been excluded because of the proximity to a sign change (dashed lines,
lower-left shaded region of Fig.\ref{fig:grid}).  One observes in
particular a spike near $x\simeq 7 \times 10^{-3}$ where the charm
distribution has its zero. Including this in a estimate of the global
accuracy would be senseless since it actually corresponds to a
divergence and the peak-value for the spike is arbitrary, depending on
the precise choice of points used to estimate the accuracy. The
question of the exact region to exclude is somewhat arbitrary, but the
choice made above seems not unreasonable in the light of
Fig.~\ref{fig:acc-fixed-dy-dlnlnQ}.

Fig.~\ref{fig:acc-fixed-dy-dlnlnQ} is useful in order to obtain a
detailed picture of the accuracy of the evolution with a given set of
parameters. To quote a single, global, number for the relative 
accuracy $\epsilon$ we make
the conservative choice of taking the largest value of $\epsilon$ that
occurs in a chosen $x$ range. We will examine a restricted range,
$x<0.7$, studying just the $g,u,d,s$ flavours, and also a wider range,
$x<0.9$ with all flavours.

\begin{figure}
  \centering
  \includegraphics[width=0.48\textwidth]{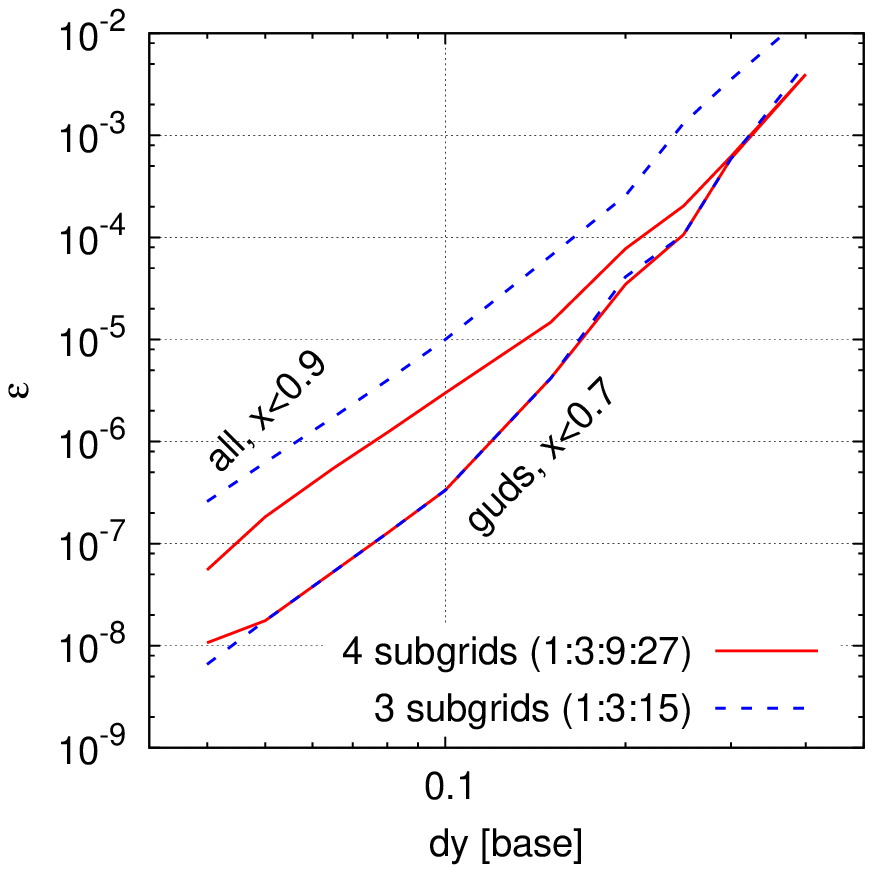}%
  \hfill
  \includegraphics[width=0.495\textwidth]{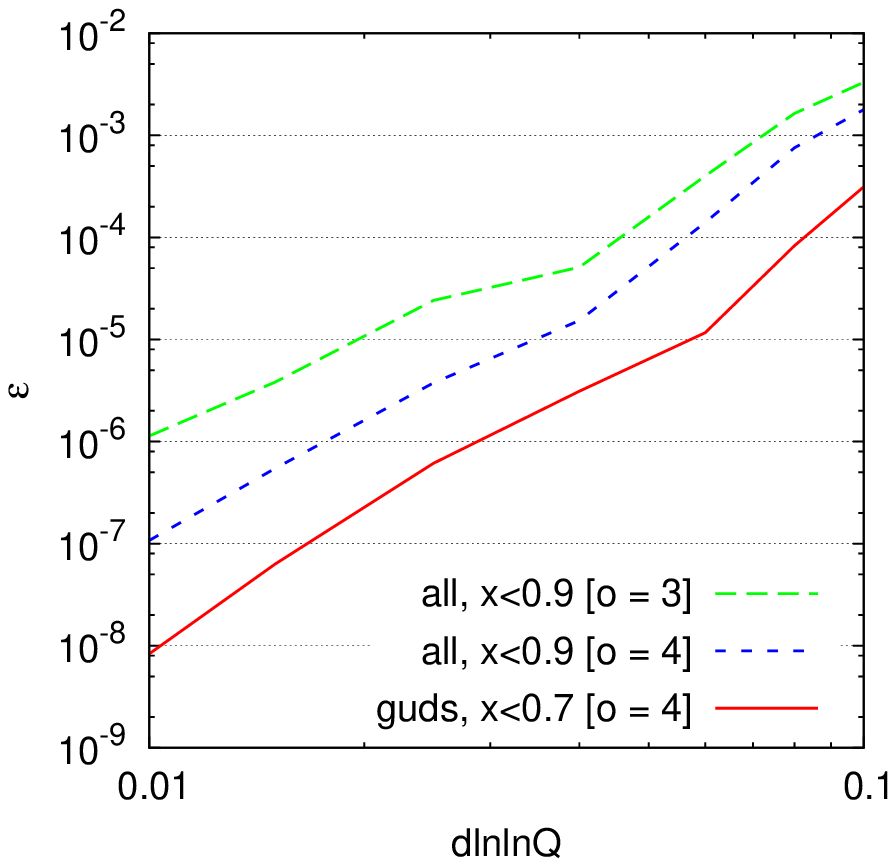}%
  \caption{Left: the (globally) worst relative accuracy $\epsilon$ as
    a function of the base $y$-grid resolution parameter, \ttt{dy} ---
    shown for two $y$-grid configurations and two $x$-ranges and
    flavour-sets.
  Right: the relative accuracy as a function of the
    resolution in $\ln \ln Q$ of the tabulation, \ttt{dlnlnQ}, for
    different $x$/flavour ranges and for different $\ttt{lnlnQ\_order}$
    values (\ttt{o}).}
  \label{fig:dy+dlnlnQ}
\end{figure}

Fig.~\ref{fig:dy+dlnlnQ} shows the effect of varying the base $\dy$
and $\dlnlnQ$ separately, while the other is fixed at the reference
value. The `guds' flavours in the $x<0.7$ range are generally better
determined, for a given set of grid parameters, than the full set of
flavours up to $x<0.9$. This is as one would expect since the
large-$x$ region is usually the hardest and the `guds' flavours are
generally somewhat smoother than the the others. Using only three $y$
subgrids worsens the situation when including the largest $x$ values,
and reducing the order in the $Q$ interpolation also adversely affects
the accuracy (also for $x<0.7$, not shown).

From Fig.~\ref{fig:dy+dlnlnQ}, we deduce that inaccuracies from the
$Q$ and $y$ parts of the grid are similar when $\dlnlnQ = \dy /4$.
This is the combination that we shall use as standard.

%......................................................................
\subsection{Timing}
\label{sec:Timing}

The time spent in \hoppet for a given analysis can expressed as
follows, according to whether or not one carries out pre-evolution:
\begin{subequations}
  \label{eq:timing}
  \begin{align}
    t_\text{no pre-ev}   &= t_s + n_\alpha t_\alpha + n_i (t_i  + n_{xQ}\, t_{xQ})\,,\\
    t_\text{with pre-ev} &= t_s + n_\alpha (t_\alpha + t_p) + n_i (t_c + n_{xQ}\,
    t_{xQ})\,,
  \end{align}
\end{subequations}
where $t_s$ is the time for setting up the splitting functions,
$n_\alpha$ is the number of different running couplings that one has,
$t_\alpha$ is the time for initialising the coupling,
$n_i$ is the number of PDF initial conditions that one wishes to
consider, $t_i$ is the time to carry out the tabulation for a single
initial condition, $n_{xQ}$ is the number of points in $x,Q$ at which
one evaluates the full set of flavours once per PDF initial condition;
in the case with pre-prepared cached evolution, $t_p$ is the time for a
preparing a cached evolution and $t_c$ is the time for performing the
cached evolution. Finally $t_{xQ}$ is the time it takes to evaluate
the PDFs at a given value of $(x,Q^2)$ once the tabulation has
been performed.

\begin{table}
  \centering
  \begin{tabular}{|ll|c|c|c|}\hline
          &&     lf95  &  ifort   & g95    \\\hline
   $t_s$ &[s]  &  0.9  &  0.66    & 2.8    \\
   $t_\alpha$ & [ms]                
               &  0.16 &  0.12    & 0.13   \\
   $t_i$ &[ms] &  37   &   38     & 330    \\
   $t_p$ &[ms] &  51   &   44     & 310    \\
   $t_c$ &[ms] &  8.8  &   9.8    & 110    \\
   $t_{xQ}$ &[$\mu$s]           
               &  2.7  &   3.1    &  25    \\
   \hline
  \end{tabular}
  \caption{Contributions to the run time in eqs.~(\ref{eq:timing})
 for    $\dy=0.2$ and 
    $\dlnlnQ=0.05$ and standard values for the other parameters 
    (on a 3.4GHz Pentium IV (D) with 2~MB cache).}
  \label{tab:timings}
\end{table}

The various contributions to the run-time are shown in
table~\ref{tab:timings} for $\dy=0.2$ and $\dlnlnQ=0.05$ (giving an
accuracy $\sim 10^{-4}$), for various compilers.  In a typical
analysis where run-times matter, such as a PDF fit, it is to be
expected that the time will be dominated by $t_c$ (or $t_i$). However,
in the typical case of global PDF fits, for which
the number of $x,Q$ points is rather large ($\gtrsim 3000$), 
 it will be $n_{xQ} t_{xQ}$ that takes the most time\footnote{
With these numbers, it is easy to check that a global fit with
$n_{xQ}\sim 3000$ and $n_i\sim 10^5$ could be completed
in less than half and hour.}. We note (with regret!) the
considerably speed advantage (almost an order
of magnitude) that is to be had with commercial
compilers.

\begin{figure}
  \centering
  \includegraphics[width=0.7\textwidth]{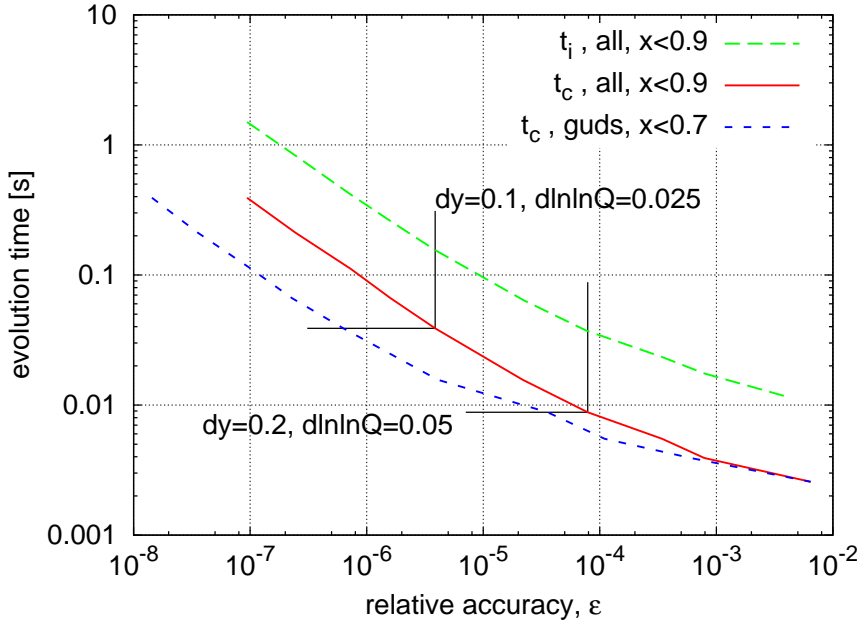}%
  \caption{relative accuracy obtained as a function of the time taken
    to perform the evolution (lf95, 3.4GHz Pentium IV (D) with 2~MB
    cache).}
  \label{fig:acc-v-time}
\end{figure}

We study $t_c$ and $t_i$ in more detail in Fig.~\ref{fig:acc-v-time},
where we relate them to the accuracy obtained from the evolution. As
one would expect, studying just the `guds' flavours for $x<0.7$ one
obtains better accuracy for a given speed than with all flavours for
$x<0.9$. Overall one can obtain $10^{-4}$ accuracy with $t_c \simeq
10^{-2}$\,s and $10^{-6}$ accuracy with $t_c \simeq 10^{-1}$\,s. In
general $t_i$ is about $4-5$ larger than $t_c$, highlighting the
advantage of the cached evolution. 

We note that the time $t_{xQ}$ for evaluating each point is
essentially independent of $\dy$ and $\dlnlnQ$. If a computation is
dominated by $t_{xQ}$, then it can be made somewhat faster by
lowering the interpolation orders, at the expense of needing a finer
grid (and so longer evolution times).

The timings shown here are roughly similar, for accuracies $\sim
10^{-4}$, to those obtained with the $N$-space code
Pegasus~\cite{Pegasus} when the number of $x,Q$ points to be
evaluated is $\order{10^3}$. For much smaller numbers of points
Pegasus becomes superior (because of the significantly smaller ratio
$t_c/t_{xQ}$), while for much larger numbers of points \hoppet becomes
better. Other NNLO evolution codes published in recent years 
\cite{Weinzierl:2002mv,coriano,Botje}
 are generally less competitive
either in terms of accuracy or speed.

To close this section, we summarise in table~\ref{tab:acc-param} the
different parameters that are relevant in determining the accuracy of
the evolution and tabulation, together with comments about the
components of the timing affected by each parameter.

\begin{table}
  \centering
  \begin{tabular}{|l|c|c|l|} \hline
    Parameter & Default & Timing impact & Notes \\ \hline
    base \ttt{dy}  &  $-$    &   $t_s, t_i, t_p, t_c$    & Default subgrids
    in ratio 1:3:9:27 \\
    \ttt{order} &  $[-6]$    &   $t_s, t_p, t_{xQ} (t_i, t_c)$ & \\ 
    \ttt{DefaultConvolutionEps} & $10^{-7}$ & $t_s$ & final acc.\
    limited by $\sim$ twice this.\\ \hline
    \ttt{du}  &  $0.1$ &  $t_i, t_p$ & immaterial if $\gtrsim 1.4\,
    \ttt{dlnlnQ}$\\
    \ttt{dlnlnQ}       &  $[\ttt{dy}/4] $ &  $t_i, t_p, t_c$ &  \\
    \ttt{lnlnQ\_order} &  $4$ &  $t_{xQ}$ &  \\ \hline
    \ttt{DefaultCouplingDt} & $0.2$ & $t_\alpha$ & default sufficient for 
$\epsilon \sim 10^{-9}$\\
    \hline
  \end{tabular}
  \caption{Parameters involved in the accuracy
of a tabulated evolution. Default values shown in square
    brackets are to be specified by hand in the F95 interface, but are
    automatically set in the simpler of the initialisation calls with
    the streamlined interface.} 
%\comment{new section for subgrids? Exact
%      v. Param}}
  \label{tab:acc-param}
\end{table}

%% Notes:
%% \begin{itemize}
%% \item Protection scheme regarding sign change: if a sign change occurs
%%   within $0.4$ of the point's $y$ value (decreasing gradually to
%%   $0.04$ at large $x$) or in neighbouring $Q$ values, then the point
%%   is discarded (for that flavour).
%% \end{itemize}
%% 
%% Comparisons to $N$-space codes (especially Vogt):
%% 
%% \begin{itemize}
%% \item $N$-space is good for a moderate (few hundred) number of $x$-$Q$
%%   points (for lots of points, better to convert it to tabulation)
%% %
%% \item For $\sim 500$ points, $N$-space is faster for the same
%%   accuracy, and its running times probably scale much more tamely when
%%   you try to go to extremely high accuracies.
%% %
%% \item Nevertheless, for reference purposes the $x$-space code can also
%%   go to high accuracies ($10^{-7}$) if you're willing to accept slow
%%   runs.
%% %
%% \item At accuracies that are acceptable for phenomenology ($10^{-3} -
%%   10^{-4}$) the $x$-space code runs nearly as fast as Vogt's $N$-space
%%   code in its lower accuracy option (which is still better accuracy
%%   than the $x$-space code). But overheads are different --- evolution
%%   is a one-off operation, accessing points is then done using
%%   interpolation.
%% %
%% \item $N$-space isn't an option if you don't have a simple analytic
%%   form for your initial parton distributions.
%% %
%% \item currently no other code has the option of the exact NNLO
%%   splitting functions.
%% \end{itemize}

%======================================================================
\section{Conclusions}

\hoppet is an $x$-space evolution code that is novel both in terms of
the accuracy and speed that it provides compared to other $x$-space
codes, and in terms of its interface, designed to provide a
straightforward and physical way of manipulating PDFs beyond the
built-in task of DGLAP evolution.

Features that might be envisaged for future releases include DIS
coefficient functions, full support for the DIS factorisation scheme,
and the addition of time-like evolution, relevant
for phenomenological fits to fragmentation functions
as in \cite{de Florian:2007hc}.  
In principle, the
information presented here is sufficient to allow a user to implement
the coefficient functions themselves, while the DIS scheme and
timelike evolution would require somewhat more knowledge of the
internals of the program. 

More ambitious possible extensions cover a wide range of physics.
Just within QCD, a general physical feature absent from mainstream PDF
evolution codes is that of evolution that includes matching with
various types of resummed calculations. Although studies in this direction
have already been performed, both for small $x$ resummations, 
as in \cite{White:2005wm}
 and for large $x$ resummation, as in \cite{Corcella:2005us},
no general public code exists which performs a matching between
resummed 
 and fixed (NLO, NNLO) order
splitting functions, either in the time-like case or in
the space-like case. 

Another interesting extension of \hoppet would be to
implement a more general mass treatment of heavy quarks.
A proper treatment of heavy quark mass effects is required to
obtain a good description of heavy flavour structure function
as measured in HERA. Although there are by now several studies
of the effect of heavy quark masses in global analysis
of PDFs \cite{Martin:2007bv,Tung:2006tb},
 which show sizable effects on predictions for
LHC observables, there does not exist right now a public code
were this General Mass heavy quark schemes are implemented.

Also of interest are non-QCD effects. Evolutions with QED radiation
have been presented in \cite{Weinzierl:2002mv,Martin:2004dh}, 
however so far no public code
exists for evolution including both QCD and electroweak (EW) effects
\cite{Ciafaloni:2000df,Ciafaloni:2005fm}.
This is of particular relevance at LHC energies, since flavour is
associated with an SU(2) charge and so soft divergences (above $M_W$)
do not cancel in the PDF evolution between real and virtual
contributions, leading one to expect
non-negligible effects in the flavour structure of the PDFs at high
scales.
The full QCD+EW evolution is a rather task complicated because of the
need to include the EW flavour couplings, including the CKM matrix,
and polarisation. For this kind of problem a code such as \hoppet
provides a good starting point, since it has a clean separation of the
numerical and flavour aspects of evolution, and verified unpolarised
and polarised evolution.

%======================================================================
\section*{Acknowledgements}

This work was initiated in the context of DIS event shape resummation
and matching studies \cite{DisResum} with Mrinal Dasgupta. Its
subsequent development into a fully-featured NNLO evolution code owes
much to Andreas Vogt's regular encouragement, and his suggestions
about features that would be useful to include for benchmark tests.
We are grateful also to Wu-Ki Tung for comments on the documentation.
This work was supported in part by grant ANR-05-JCJC-0046-01 from the
French Agence Nationale de la Recherche. 

%======================================================================
\appendix

%======================================================================
\section{Example programs}
\label{sec:example_program}

\subsection{General interface}

The program below generates a subset of table~15 of the NNLO benchmark
evolution in the second reference of~\cite{Benchmarks}. It is to be
found (in a slightly more commented form) in
\ttt{example\_f90/tabulation\_example.f90}. Compilation instructions
are to be found in the \ttt{README} file in the main directory of the
release. A program that has the same functionality but in F77, using
the streamlined interface of Sect.~\ref{sec:vanilla}, is to be found as
\ttt{example\_f77/tabulation\_example.f}.

\begin{lstlisting}
program tabulation_example
  use hoppet_v1
  implicit none
  real(dp)         :: dy, ymax, quark_masses(4:6)
  integer          :: order, nloop, ix
  type(grid_def)   :: grid, gdarray(4) ! holds information about the grid
  type(dglap_holder)     :: dh         ! holds the splitting functions
  type(pdf_table)        :: table      ! holds the PDF tabulation
  type(running_coupling) :: coupling  
  real(dp), pointer      :: pdf0(:,:)  ! holds the initial pdf
  real(dp)               :: Q0, Q, pdf_at_xQ(-6:6)
  real(dp), parameter :: heralhc_xvals(9) = &
       & (/1e-5_dp,1e-4_dp,1e-3_dp,1e-2_dp,0.1_dp,0.3_dp,0.5_dp,0.7_dp,0.9_dp/)

  ! set up parameters for grid
  order = -6
  ymax  = 12.0_dp
  dy    = 0.1_dp
  
  ! set up the grid itself -- we use 4 nested subgrids
  call InitGridDef(gdarray(4),dy/27.0_dp, 0.2_dp, order=order)
  call InitGridDef(gdarray(3),dy/9.0_dp,  0.5_dp, order=order)
  call InitGridDef(gdarray(2),dy/3.0_dp,  2.0_dp, order=order)
  call InitGridDef(gdarray(1),dy,          ymax , order=order)
  call InitGridDef(grid,gdarray(1:4),locked=.true.)

  ! initialise the splitting-function holder
  nloop = 3
  call InitDglapHolder(grid,dh,factscheme=factscheme_MSbar,&
       &                      nloop=nloop,nflo=3,nfhi=6)

  ! initialise a PDF from the function below (must be contained,
  ! in a "used" module, or with an explicitly defined interface)
  call AllocPDF(grid, pdf0)
  pdf0 = unpolarized_dummy_pdf(xValues(grid))
  Q0 = sqrt(2.0_dp)  ! the initial scale

  ! allocate and initialise the running coupling with a given
  ! set of quark masses (NB: charm mass just above Q0).
  quark_masses(4:6) = (/1.414213563_dp, 4.5_dp, 175.0_dp/)
  call InitRunningCoupling(coupling,alfas=0.35_dp,Q=Q0,nloop=nloop,&
       &                   quark_masses = quark_masses)

  ! create the tables that will contain our copy of the user's pdf
  ! as well as the convolutions with the pdf.
  call AllocPdfTable(grid, table, Qmin=1.0_dp, Qmax=10000.0_dp, & 
       & dlnlnQ = dy/4.0_dp, freeze_at_Qmin=.true.)
  ! add information about the nf transitions to the table (improves
  ! interpolation quality)
  call AddNfInfoToPdfTable(table,coupling)

  ! create the tabulation based on the evolution of pdf0 from scale Q0
  call EvolvePdfTable(table, Q0, pdf0, dh, coupling, nloop=nloop)
  ! alternatively "pre-evolve" so that subsequent evolutions are faster
  !call PreEvolvePdfTable(table, Q0, dh, coupling)
  !call EvolvePdfTable(table,pdf0)

  ! get the value of the tabulation at some point
  Q = 100.0_dp
  write(6,'(a,f8.3,a)') "           Evaluating PDFs at Q = ",Q," GeV"
  write(6,'(a5,2a12,a14,a10,a12)') "x",&
       & "u-ubar","d-dbar","2(ubr+dbr)","c+cbar","gluon"
  do ix = 1, size(heralhc_xvals)
     call EvalPdfTable_xQ(table,heralhc_xvals(ix),Q,pdf_at_xQ)
     write(6,'(es7.1,5es12.4)') heralhc_xvals(ix), &
          &  pdf_at_xQ(2)-pdf_at_xQ(-2),  pdf_at_xQ(1)-pdf_at_xQ(-1), &
          &  2*(pdf_at_xQ(-1)+pdf_at_xQ(-2)), (pdf_at_xQ(-4)+pdf_at_xQ(4)), &
          &  pdf_at_xQ(0)
  end do
  
  ! some cleaning up (not strictly speaking needed, but illustrative)
  call Delete(table); call Delete(pdf0); call Delete(dh)
  call Delete(coupling);  call Delete(grid)

contains 
  !======================================================================
  !! The dummy PDF suggested by Vogt as the initial condition for the 
  !! unpolarized evolution (as used in hep-ph/0511119).
  function unpolarized_dummy_pdf(xvals) result(pdf)
    real(dp), intent(in) :: xvals(:)
    real(dp)             :: pdf(size(xvals),-6:7) ! note upper bound!
    real(dp) :: uv(size(xvals)), dv(size(xvals))
    real(dp) :: ubar(size(xvals)), dbar(size(xvals))
    !---------------------
    real(dp), parameter :: N_g = 1.7_dp, N_ls = 0.387975_dp
    real(dp), parameter :: N_uv=5.107200_dp, N_dv = 3.064320_dp
    real(dp), parameter :: N_db = half*N_ls
  
    pdf = zero

    !-- remember that these are all xvals*q(xvals)
    uv = N_uv * xvals**0.8_dp * (1-xvals)**3
    dv = N_dv * xvals**0.8_dp * (1-xvals)**4
    dbar = N_db * xvals**(-0.1_dp) * (1-xvals)**6
    ubar = dbar * (1-xvals)

    ! labels iflv_g, etc., come from the hoppet_v1 module
    pdf(:, iflv_g) = N_g * xvals**(-0.1_dp) * (1-xvals)**5
    pdf(:,-iflv_s) = 0.2_dp*(dbar + ubar)
    pdf(:, iflv_s) = pdf(:,-iflv_s)
    pdf(:, iflv_u) = uv + ubar
    pdf(:,-iflv_u) = ubar
    pdf(:, iflv_d) = dv + dbar
    pdf(:,-iflv_d) = dbar
  end function unpolarized_dummy_pdf

end program tabulation_example
\end{lstlisting}

\noindent The expected output from the program is:
{\small
\begin{lstlisting}
           Evaluating PDFs at Q =  100.000 GeV
    x      u-ubar      d-dbar    2(ubr+dbr)    c+cbar       gluon
1.0E-05  3.1907E-03  1.9532E-03  3.4732E+01  1.5875E+01  2.2012E+02
1.0E-04  1.4023E-02  8.2749E-03  1.5617E+01  6.7244E+00  8.8804E+01
1.0E-03  6.0019E-02  3.4519E-02  6.4173E+00  2.4494E+00  3.0404E+01
1.0E-02  2.3244E-01  1.3000E-01  2.2778E+00  6.6746E-01  7.7912E+00
1.0E-01  5.4993E-01  2.7035E-01  3.8526E-01  6.4466E-02  8.5266E-01
3.0E-01  3.4622E-01  1.2833E-01  3.4600E-02  4.0134E-03  7.8898E-02
5.0E-01  1.1868E-01  3.0811E-02  2.3198E-03  2.3752E-04  7.6398E-03
7.0E-01  1.9486E-02  2.9901E-03  5.2352E-05  5.6038E-06  3.7080E-04
9.0E-01  3.3522E-04  1.6933E-05  2.5735E-08  4.3368E-09  1.1721E-06
\end{lstlisting}
}

\noindent The file
\ttt{example\_f90/tabulation\_example.default\_output} contains a copy
of these results, so as to allow easy comparison. The numbers
correspond to evolution with variable flavour number, $\mu_F = \mu_R$,
and the parametrised versions of the NNLO splitting functions and mass
threshold terms.
The reader may verify that they correspond to those given in the top
panel of table~15 of the second reference of \cite{Benchmarks} ($\mu_F
= \mu_R$).

\subsection{Streamlined interface}

The program below generates exactly the same output
as the previous example program, but this time
using the streamlined interface introduced
in Sect.~\ref{sec:vanilla}. It is to be
found in
\ttt{example\_f90/tabulation\_example\_streamlined.f90}. 
A program with same interface and same output 
but in F77, is to be found in
\ttt{example\_f77/tabulation\_example.f}.

\begin{lstlisting}
program tabulation_example_streamlined
  use hoppet_v1
  !! if using LHAPDF, rename a couple of hoppet functions which
  !! would otherwise conflict with LHAPDF 
  !use hoppet_v1, EvolvePDF_hoppet => EvolvePDF, InitPDF_hoppet => InitPDF
  implicit none
  real(dp) :: dy, ymax, dlnlnQ, Qmin, Qmax, muR_Q
  real(dp) :: asQ, Q0alphas, Q0pdf
  real(dp) :: mc,mb,mt
  integer  :: order, nloop
  !! holds information about the grid
  type(grid_def) :: grid, gdarray(4)
  !! hold results at some x, Q
  real(dp) :: Q, pdf_at_xQ(-6:6)
  real(dp), parameter :: heralhc_xvals(9) = &
       & (/1e-5_dp,1e-4_dp,1e-3_dp,1e-2_dp,0.1_dp,0.3_dp,0.5_dp,0.7_dp,0.9_dp/)
  integer  :: ix

  ! set up parameters for grid
  order = -6
  ymax  = 12.0_dp
  dy    = 0.1_dp

  ! set up the grid itself -- we use 4 nested subgrids
  call InitGridDef(gdarray(4),dy/27.0_dp,0.2_dp, order=order)
  call InitGridDef(gdarray(3),dy/9.0_dp,0.5_dp, order=order)
  call InitGridDef(gdarray(2),dy/3.0_dp,2.0_dp, order=order)
  call InitGridDef(gdarray(1),dy,       ymax  ,order=order)
  call InitGridDef(grid,gdarray(1:4),locked=.true.)

  ! Streamlined initialisation
  Qmin=1_dp
  Qmax=28000_dp
  dlnlnQ = dy/4.0_dp
  nloop = 3
  call hoppetStartExtended(ymax,dy,Qmin,Qmax,dlnlnQ,nloop,&
       &         order,factscheme_MSbar)
    
  ! Set heavy flavour scheme
  mc = 1.414213563_dp
  mb = 4.5_dp
  mt = 175.0_dp
  call hoppetSetVFN(mc, mb, mt)
  
    ! Set parameters of running coupling
  asQ = 0.35_dp
  Q0alphas = sqrt(2.0_dp)
  muR_Q = 1.0_dp
  Q0pdf = sqrt(2.0_dp) ! The initial evolution scale
  ! Normal evolution
  call hoppetEvolve(asQ, Q0alphas, nloop,muR_Q,&
       &       LHAsub, Q0pdf)

  ! Uncomment to perform  cached evolution
  ! call hoppetPreEvolve(asQ, Q0alphas, nloop,muR_Q,Q0pdf) 
  ! call hoppetCachedEvolve(LHAsub)
    
  ! get the value of the tabulation at some point
  Q = 100.0_dp
  write(6,'(a)')
  write(6,'(a,f8.3,a)') "           Evaluating PDFs at Q = ",Q," GeV"
  write(6,'(a5,2a12,a14,a10,a12)') "x",&
       & "u-ubar","d-dbar","2(ubr+dbr)","c+cbar","gluon"
  do ix = 1, size(heralhc_xvals)
     call hoppetEval(heralhc_xvals(ix),Q,pdf_at_xQ)
     write(6,'(es7.1,5es12.4)') heralhc_xvals(ix), &
          &  pdf_at_xQ(2)-pdf_at_xQ(-2), &
          &  pdf_at_xQ(1)-pdf_at_xQ(-1), &
          &  2*(pdf_at_xQ(-1)+pdf_at_xQ(-2)), &
          &  (pdf_at_xQ(-4)+pdf_at_xQ(4)), &
          &  pdf_at_xQ(0)
  end do
  
contains 
 
  subroutine LHAsub(x,Q,pdf)
    ! Same as in the previous example program
  end subroutine LHAsub

end program tabulation_example_streamlined
\end{lstlisting}

\subsection{Accessing tables}

As has been mentioned in Sect.~\ref{sec:acc_table}, 
there are several ways to define, initialise
and access a user defined table, for example 
as in the example below:
\begin{lstlisting}
module external_table_module ! common location for your table
   use hoppet_v1
   implicit none
   type(pdf_table) :: table
   type(dglap_holder) :: dh
   type(running_coupling) :: coupling
   type(grid_def) :: grid   ! Optional
end module external_table_module
\end{lstlisting}
Note that one might need to include in the common module objects
like \texttt{dh} or \ttt{coupling}, since these are
required in other procedures:
\begin{lstlisting}

subroutine A
   use external_table_module
   call PreEvolvePdfTable(table, Q0, dh, coupling)
end subroutine A

subroutine B
   use external_table_module
   ...
   call EvolvePdfTable(table,pdf0)
end subroutine B

subroutine C
   use external_table_module
   ...
   call EvalPdfTable_xQ(table, x, Q, pdf)
end subroutine C

\end{lstlisting}

A more detailed description of the above technique 
can be found in a 
second example program, called \ttt{tabulation\_example\_2.f90},
which is available at the \ttt{example\_f90} 
directory. It generates exactly
the same output as the previous example program, however it
provides an example of how to access a table from
external procedures, as explained in Sect.~\ref{sec:acc_table}.

%===================================================================

%==========================================================
% In future releases of the documentation

%\section{Public objects}
%In this Appendix we list some of the
%most important public objects that are
%accessible through the common {\tt hoppet\_v1} module. For
%simplicity we describe here only the public procedures and
%not the public variables.

%\begin{table}
%\scriptsize
%\begin{center}
%\begin{tabular}{|c|c|c|c|}
%\hline
%{\bf module}  & {\bf procedure}  &  {\bf arguments}  & {\bf description} \\
%\hline
%\ttt{pdf\_tabulate}  &   & & \\
%  &  \ttt{AllocPdfTable} & &\\
%\hline
%\ttt{pdf\_representation} && &\\
%\hline
%\end{tabular}
%\end{center}
%\end{table}

%=======================================================

\section{HOPPET reference guide}
\label{sec:refguide}

In this section we present the \hoppet reference guide, a 
summary of the most important modules in the package with
the corresponding description, both the the streamlined interface,
Table \ref{tab:streamlined} and for the general interface,
Table  \ref{tab:general}.

\begin{table}
\begin{center}
{\bf \large STREAMLINED INTERFACE}
\\
\vspace{0.2cm}
\begin{tabular}{|c|c|}
\hline
\bf METHOD  & \bf DESCRIPTION \\
\hline
\bf Initialisation & \\
\hline
\begin{lstlisting}
 hoppetStart(dy,nloop)
\end{lstlisting} 
& \scriptsize
Sets up a compound grid with
spacing in $\ln 1/x$ of \ttt{dy} at small $x$,\\ &
\fn~  extending to $y = 12$ and numerical
order $\ttt=-5$. \\ & \fn~  The $Q$ range for the tabulation will be $1\GeV <
Q<28 \TeV$, \\
& \fn~ \ttt{dlnlnQ=dy/4} and the factorisation scheme is ${\overline{\rm MS}}$\\
\hline
\begin{lstlisting}
 hoppetStartExtended(ymax,dy,Qmin,
 Qmax,dlnlnQ,nloop,order,factscheme)
\end{lstlisting} & \fn
  ~More general initialisation \\
\hline 
\begin{lstlisting}
 hoppetSetFFN(fixed_nf)
 hoppetSetVFN(mc, mb, mt)
\end{lstlisting} &
\fn ~Set heavy flavour scheme
\\
\hline
\begin{lstlisting}
  alphas = hoppetAlphaS(Q)
\end{lstlisting} &
\fn~Accessing the coupling \\
\hline&\\[-0.5em]
\bf Normal evolution & \\
\hline
\texttt{\footnotesize
 hoppetEvolve(asQ,Q0alphas,}
& \fn\ PDF evolution: specifies the coupling \ttt{asQ} at a 
scale \ttt{Q0alphas}, \\
 \texttt{\footnotesize nloop,muR\_Q,LHAsub,Q0pdf)}&
 \fn the number of loops for  evol., \ttt{nloop}, \\
&\fn the ratio (\ttt{muR\_Q}) of ren. to fact. scales. \\ 
&  \fn the name of a subroutine \ttt{LHAsub} with an LHAPDF-like interface \\
&  \fn and the scale
\ttt{Q0pdf} at which one starts the PDF evolution \\
&  \fn Note:  \ttt{LHAsub} only called at scale \ttt{Q0pdf}\\
\hline
\begin{lstlisting}
 hoppetEval(x,Q,f)
\end{lstlisting} &
\fn On return, \ttt{f(-6:6)} contains all flavours of the PDF set\\ 
%The PDF at a given value of  is obtained \\
& \fn ~
(multiplied by $x$) at the given $\ttt{x}$ and $\ttt{Q}$ values \\
\hline&\\[-0.5em]
\bf Cached evolution & \\
\hline
\begin{lstlisting}
 hoppetPreEvolve(asQ,Q0alphas, 
          nloop, muR_Q, Q0pdf)
\end{lstlisting} & \fn ~
 Preparation of the cached evolution\\
\hline
\begin{lstlisting}
 hoppetCachedEvolve(LHAsub)
\end{lstlisting} &
\fn  Perform cached evolution with the initial condition\\
& \fn at \ttt{Q0pdf} from a routine \ttt{LHAsub} 
with LHAPDF-like interface\\
&  \fn Notice  \ttt{LHAsub} only called at scale \ttt{Q0pdf}\\
\hline
\begin{lstlisting}
 hoppetEval(x,Q,f)
\end{lstlisting} &
\fn On return, \ttt{f(-6:6)} contains all flavours of the PDF set\\ 
%The PDF at a given value of  is obtained \\
& \fn ~
(multiplied by $x$) at the given $\ttt{x}$ and $\ttt{Q}$ values \\
& \fn [as for normal evolution]\\
\hline
\end{tabular}
\end{center}
\caption{\label{tab:streamlined} Reference guide for the streamlined
  interface. Note that \hoppet should be supplied with and returns parton
  densities multiplied by $x$.}
\end{table}

%%%%%%%%%%%%%%%%%%%%%%%%%%%%%%%%%%%%%%%%%%%%%

% \begin{table}
% 
% 
% \\
% 

\begin{table}
\scriptsize
\begin{center}
{\bf \large GENERAL INTERFACE} \\
\vspace{0.5cm} 
%\scalebox{0.01}{
\begin{tabular}{|c|c|}
\hline
\bf TYPES  & \bf DESCRIPTION \\
\hline
\texttt{type(grid\_def) :: grid} & $x-$space grid definition \\ 
\hline
\texttt{real(dp), pointer :: gluon(:)} & Holds a 
`grid quantity' (\eg gluon PDF) \\
\hline
\texttt{real(dp), pointer :: PDFset(:,:)} & 
Grid representation of a (13-flavour) PDF set \\
\hline
\texttt{type(grid\_conv) :: Pgg} 
 & Convolution operator ({\it i.e.} splitting function) \\ 
\hline
\texttt{ type(split\_mat) :: Pmat}
 & Splitting matrix (with full flavour structure)\\ 
\hline
 \texttt{type(mass\_threshold\_mat) :: MTM\_NNLO}
 & Heavy quark mass-threshold matrix\\ 
\hline
\texttt{ type(dglap\_holder) :: dglap\_h} & DGLAP holder (\ie all
splitting and mass-threshold matrices) \\
\hline
\texttt{type(running\_coupling) :: coupling} & Running coupling \\
\hline
\texttt{type(evln\_operator) :: evop} & Evolution operator (linked
list of split \& mass-threshold matrices)\\
\hline
\texttt{type(pdf\_table) :: table} & PDF set tabulated in $x$ \& $Q$\\
\hline
\end{tabular}
\vspace{0.7cm}\\
\begin{tabular}{|c|c|}
\hline
\bf METHOD  & \bf DESCRIPTION \\
\hline &\\[-0.5em]
\bf Initialisation & \\[0.3em]
\hline
\texttt{
 InitGridDef(grid,dy=0.1\_dp,ymax=10.0\_dp,order=3)
}
 &
~ ~Initialise a grid definition \\
\hline
\texttt{
InitGridDef(grid,subgrids(:),locked=.true.)
}
 &
~ ~Combine subgrids into a single grid \\
\hline
\texttt{
 AllocGridQuant(grid,gluon)
} 
 &
Allocate memory for a grid quantity (\eg gluon PDF) \\
\hline
\texttt{
 InitGridQuant(grid,gluon,example\_gluon\_fn)
} 
 &
~ ~Initialisation of a grid quantity (\eg gluon PDF) \\
\hline
\texttt{
 AllocPDF(grid,pdfset)
} 
 &
Allocate memory for a 13-flavour PDF set \\
\hline
\texttt{InitPDF\_LHAPDF(grid,pdfset,LHAsub,Q)} 
 &
   Initialisation of a (13-flavour) PDF set from LHAPDF \\
&  type routine. Note: it only calls \ttt{LHAsub} at the scale \ttt{Q}\\
\hline
\texttt{
InitGridConv(grid,Pgg,Pgg\_func)
}  (*)
 &
~ ~Initialisation of a convolution operator \\
\hline
\texttt{
InitDglapHolder(grid, dglap\_h, factscheme, nloop)
}  (*)
&
~ ~Initialisation of a {\tt dglap\_holder} type \\
\hline
\texttt{
 InitRunningCoupling(coupling [, alfas] [, Q]} 
&
~ ~Initialisation of a {\tt running\_coupling} type \\
\texttt{[, nloop] [, fixnf] [, quark\_masses] )}  (*)&
\\
\hline
\texttt{
 AllocPdfTable(grid, table, Qmin, Qmax}  
&
~ ~Allocate space for a {\tt pdf\_table} type \\
\texttt{[, dlnlnQ ] [, lnlnQ\_order ] [, freeze\_at\_Qmin] )
} (*) & \\
\hline &\\[-0.5em]
\bf Evaluation \& manipulation & \\[0.3em]
\hline
\texttt{
EvalGridQuant(grid,gluon,y)
}
 & ~~Evaluation of a grid quantity at $y=\ln 1/x$\\
\hline
\texttt{Pgg.conv.gluon} & Convolution of a splitting function with a
(1-flav) PDF  \\
\hline
\texttt{Pmat.conv.PDFset} & Convolution of a splitting matrix
with a PDF set  \\
\hline
\texttt{
SetToConvolution(Pab,Pac,Pcb) 
}
 & ~~Convolution of splitting functions, \ttt{Pab = Pac Pcb} \\
\hline
\texttt{
 EvalPdfTable\_yQ(table, y, Q, pdf)
}
 & Evaluate the 13 flavours, \ttt{pdf(-6:6)}, of a tabulated \\
 & PDF set at $y$ and $Q$\\
\hline&\\[-0.5em]
\bf Evolution  & \\[0.3em]
\hline 
\texttt{
 CopyHumanPdfToEvln(nf\_lcl, pdf\_human, pdf\_evln)
} & ~~
Transform PDF set from {\tt human} to {\tt evln}
representation \\
\hline
\texttt{
GetPdfRep(pdfset)
} & ~~
Check PDF set representation \\
\hline
\texttt{
 EvolvePDF(dglap\_h, initial\_pdfset, coupling,}
 & ~~
Evolution of an initial condition for a PDF set \\
\texttt{ Q\_init,Q\_end ,[, muR\_Q] [, nloop] [,du])} 
& at {\tt Q\_init} to  {\tt Q\_end} \\
\hline
\texttt{
 EvolvePdfTable(table, Q0, initial\_pdfset, dglap\_h,} 
 & ~~ 
Fill a table starting from an initial condition,\\
\texttt{ coupling  [, muR\_Q] [, nloop] [, untie\_nf])} 
& \ttt{initial\_pdfset}, at the scale {\tt Q0} \\ 
\hline
\end{tabular} \\
%}
%\vspace{0.7cm}
\end{center}
\caption{\label{tab:general}
Reference guide for the general interface.
The upper table describes the main
derived types defined in \hoppet.
The lower table summarises some of the most relevant methods. Note that arguments
between \texttt{[...]} are optional. For grid quantities and PDF sets
the user must explicitly make memory allocation calls; in other cases
(marked with a (*)), initialisation routines automatically
allocate the memory. For all types, allocated memory may be freed with
a call to the \ttt{Delete(...)} subroutine. }
\end{table}

%------------------------------------------------------

\section{Initialisation of grid quantities}
\label{sec:gridinit}
As has been mentioned in
Sect.~\ref{sec:xspc},  
there exist several ways of setting a grid quantity. In this
Appendix we describe the most important methods for initialising
a grid quantity, which we take to be a parton distribution.

There are a number of ways of setting a grid quantity. Suppose we have
a subroutine
\begin{lstlisting}
  subroutine example_gluon(y,g)
    use types                !! defines "dp" (double precision) kind
    implicit none
    real(dp), intent(in)  :: y
    real(dp), intent(out) :: g
    real(dp) :: x
    
    x = exp(-y)
    g = 1.7_dp * x**(-0.1_dp) * (1-x)**5 
  end subroutine example_gluon
\end{lstlisting}
Then we can call
\begin{lstlisting}
call InitGridQuantSub(grid,gluon,example_gluon)
\end{lstlisting}
to initialise \texttt{gluon} with a representation of the return value
from the subroutine \texttt{example\_gluon}.

An alternative way is to
make use of functions \texttt{xValues} or \texttt{yValues} that
respectively return the $x$ or $y$ values of all points on the grid:
\begin{lstlisting}
  real(dp), pointer :: gluon,xvals
  call AllocGridQuant(grid,gluon)
  call AllocGridQuant(grid,xvals)
  xvals = xValues(grid)
  gluon = 1.7_dp * xvals**(-0.1_dp) * (1-xvals)**5 
  deallocate(xvals)
\end{lstlisting}
Though more laborious insofar as one has to worry about some extra
allocation and deallocation, it has the advantage that one no longer
has to write a separate subroutine.

Finally, there is an option to initialise a multi-flavour
PDF grid with a subroutine with the same format as
\ttt{evolvePDF} from the \ttt{LHAPDF} library. This
option works as follows:

\begin{lstlisting}
real(dp),pointer :: pdf_set(:,:)
real(dp)         :: y,Q
real(dp)         :: pdf_at_y(-6:6)

Q=2   ! Initial scale

! Initialise with LHAPDF-like routine
call AllocPDF(grid,pdf_set)
call InitGridQuantLHAPDF(grid, pdf_set, LHAsub, Q)

! Evaluate the multi-flavor pdf at y
pdf_at_y = EvalGridQuant(grid,pdf_set(:,-6:6),y)

\end{lstlisting}
where an example of  the \ttt{LHAsub} subroutine can be found in
Sect.~\ref{sec:human-rep}

\section{NNLO splitting functions}
\label{sect:nnlo}
A remark concerning NNLO splitting functions: the exact NNLO splitting
functions derived by Moch, Vermaseren and Vogt
\cite{NNLO-NS,NNLO-singlet} involve long (multi-page) expressions in
terms of harmonic polylogarithms of up to weight 4. Very conveniently,
refs.~\cite{NNLO-NS,NNLO-singlet} provide the expressions directly in
terms of Fortran code.
The harmonic polylogarithms can be evaluated using the \ttt{hplog}
package of Gehrmann and Remiddi \cite{FortranPolyLog}, a copy of which
is included with the \hoppet package. 

The initial integrations needed to create the \ttt{split\_mat} objects
for the exact NNLO splitting functions for the full range of $n_f$
take of the order of minutes.  Since currently there is no option of
storing the splitting matrices in a file, this can be a bit
bothersome. So instead, by default, the program uses the approximate,
parametrised NNLO splitting functions also provided in
\cite{NNLO-NS,NNLO-singlet}. The parametrised splitting functions are
guaranteed to be accurate to within $0.1\%$ --- in practice since they
come in relatively suppressed by two powers of $\as$, the impact on
the evolution tends to be of the order of a  $10^{-5}$ relative effect
\cite{Benchmarks}.

The user can choose whether to obtain the exact or parametrised NNLO
splitting functions using the following calls (to be made before
initialising the splitting matrices)
\begin{lstlisting}
  integer :: splitting_variant
  call dglap_Set_nnlo_splitting(splitting_variant)
\end{lstlisting}
with the following variants defined (as integer parameters) in the
module \ttt{dglap\_choices}:
\begin{lstlisting}
  nnlo_splitting_exact
  nnlo_splitting_param                   [default]
  nnlo_splitting_Nfitav
  nnlo_splitting_Nfiterr1
  nnlo_splitting_Nfiterr2
\end{lstlisting}
The last 3 are the parametrisations based on fits to reduced moment
information carried out in \cite{vanNeerven:1999ca,vanNeerven:2000uj}.
Though at the time they represented a valuable (and much used) step on
the way to full NNLO results, nowadays their interest is mainly
historical.

Note that only for the \ttt{nnlo\_splitting\_exact} can the colour
constants be varied (with the caveat about $d^{abc}d_{abc}$ , as is
described in Sect.~\ref{sec:qcd}). For the other options the
NNLO splitting functions have been computed using the QCD values
for the colour factors.

%======================================================================
\section{Useful tips on Fortran~95}
\label{sec:f95appendix}

As Fortran~95's use in high-energy physics is not as widespread as
that of other languages such as Fortran~77 and C++, it is useful to
summarise some key novelties compared to Fortran~77, as well as some
points that might otherwise cause confusion. For further information
the reader is referred both to books about the language such as
\cite{F95Explained} and to web resources~\cite{F95WebResources}.

\paragraph{Free form.}  Most of the code in the \hoppet package is in
free-form. The standard extension for free-form form files is
\ttt{.f90}. There is no requirement to leave 6 blank spaces before
every line and lines can consist of up to 132 characters. The other
main difference relative to f77 fixed form is that to continue a line
one must append an append an ampersand, \ttt{\&}, to the line to be
continued. One may optionally include an ampersand as the first
non-space character of the continuation line.

For readability, many of the subprogram names in this documentation
are written with capitals at the start of each word. Note however that
free-form Fortran~95, like its fixed-form predecessors, is case
insensitive.

\paragraph{Modules, and features relating to arrays.} Fortran~95
allows one to package variables and subroutines into modules
\begin{lstlisting}
module test_module
  implicit none
  integer :: some_integer
contains
  subroutine print_array(array)
    integer, intent(in) :: array(:) ! size is known, first element is 1
                                    ! intent(in) == array will not be changed
    integer             :: i, n
    n = size(array)
    do i = 1, n
      print *, i, array(i)
    end do
  end subroutine hello_world
end module test_module
\end{lstlisting}
The variable \texttt{some\_integer} and the subroutine
\texttt{print\_array} are invisible to other routines unless they
explicitly \texttt{use} the module as in the following example:
\begin{lstlisting}
program test_program
  use test_module
  implicit none
  integer :: array1(5), array2(-2:2)
  integer :: i
  
  some_integer = 5   ! set the variable in test_module
  array1       = 0   ! set all elements of array1 to zero
  array2(-2:0) = 99  ! set elements 1..3 of array2 to equal to 3.
  array2(1:2)  = 2*array2(-1:0)  ! elements -2..0 equal twice elements -1..0

  print *, "Printing array 1"
  call print_array(array1)
  print *, "Printing array 2"
  call print_array(array2)
end program test_program
\end{lstlisting}
Constants can be assigned to arrays (\texttt{array1}) or array
subsections (\texttt{array2(-2:0)}), arrays can be assigned to arrays
of the same size (as is done for \texttt{array2(-2:0)}) and
mathematical operations apply to each element of the array (as with
the multiplication by 2).

When arrays are passed to function or subroutine that is defined in a
\texttt{use}d module, information about the size of the array is
passed along with the array itself. Note however that information
about the lower bound is \emph{not} passed, so that for both
\texttt{array1} and \texttt{array2}, \texttt{print\_array} will see
arrays whose valid indices will run from $1\ldots5$. Thus the output
from the program will be
\begin{lstlisting}
 Printing array 1
 1 0
 2 0
 3 0
 4 0
 5 0
 Printing array 2
 1 99
 2 99
 3 99
 4 198
 5 198
\end{lstlisting}
If \texttt{print\_array} wants \texttt{array} to have a different lower
bound it must specify it in the declaration, for example
\begin{lstlisting}
  integer, intent(in) :: array(-2:) ! size is known, first element is -2
\end{lstlisting}
While it may initially seem bizarre, there are good reasons for such
behaviour (for example in allowing a subroutine to manipulate multiple
arrays of the same size without having to worry about whether they all
have the same lower bounds).

\paragraph{Dynamic memory allocation, pointers.} One of the major
additions of f95 compared to f77 is that of dynamic memory allocation,
for example with pointers
\begin{lstlisting}
  integer, pointer :: dynamic_array(:)
  allocate(dynamic_array(-6:6))
  ! .. work with it ..
  deallocate(dynamic_array)
\end{lstlisting}
This is fundamental to our ability to decide parameters of the PDF
grid(s) at run-time. Pointers can be passed as arguments to subprograms.
If the subprogram does not specify the \texttt{pointer} attribute for
the dummy argument
\begin{lstlisting}
subroutine xyz(dummy_array)
  integer, intent(in) :: dummy_array(:)
\end{lstlisting}
then everything behaves as if the argument were a normal array (\eg
the default lower bound is $1$). Alternatively the subroutine can
specify that it expects a pointer argument
\begin{lstlisting}
subroutine xyz(dummy_pointer_array)
  integer, pointer :: dummy_pointer_array(:)
\end{lstlisting}
In this case the subroutine has the freedom to allocate and deallocate
the array. Note also that because a pointer to the full array
information is being passed, the lower bound of \texttt{dummy\_pointer\_array}
is now the same as in the calling routine. Though this sounds like a
technicality, it is important because a corollary it that a subroutine
can allocate a dummy pointer array with bounds that are passed back to
the calling subroutine (we need this for the flavour dimension of
PDFs, whose lower bound is most naturally $-6$). 

Note that in contrast to \ttt{C}/\ttt{C++} pointers, F95 pointers do
not explicitly need to be dereferenced --- in this respect they are
more like \ttt{C++} \emph{references}. To associate a pointer with an
object, one uses the \ttt{=>} syntax:
\begin{lstlisting}
  integer, target  :: target_object(10)
  integer, pointer :: pointer_to_object(:)
  pointer_to_object => target_object
  pointer_to_object(1:10) = 0        ! sets target_object(1:10)
\end{lstlisting}
One notes that the object that was pointed to had the \ttt{target}
attribute --- this is mandatory (unless the object is itself a
pointer).

\paragraph{Derived types.} Another feature of F95 that has been
heavily used is that of derived types (analogous to C's
\texttt{struct}):
\begin{lstlisting}
  type pair
    integer first, second
  end type pair 
\end{lstlisting}
Variables of this type can then be created and used as follows
\begin{lstlisting}
  type(pair) :: pair_object, another_pair_object
  pair_object%first  = 1
  pair_object%second = 2
  another_pair_object = pair_object
  print *, another_pair_object%second
\end{lstlisting}
where one sees that the entirety of the object can be copied with the
assignment (\texttt{=}) operator. Note that many of the derived types
used in \hoppet contain pointers and when such a derived type object
is copied, the copy's pointer just points to the same memory as the
original object's pointer. This is sometimes what you want, but on
other occasions will give unexpected behaviour: for example splitting
function types are derived types containing pointers, so when you
assign one splitting function object to another, they end up referring
to the same memory, so if you multiply one of them by a constant, the
other one will also be modified.

\paragraph{Operator overloading} While assignment behaves more or less
as expected by default with derived types (it can actually be modified
if one wants to), other operators do not have default definitions. So
if one wants to define, say, a multiplication of objects one may
associate a function with a given operator, using an interface block:
\begin{lstlisting}
module test_module
  interface operator(*)        ! provide access to dot_pairs through 
    module procedure dot_pairs ! the normal multiplication symbol
  end interface 
  interface operator(.dot.)    ! provide access to dot_pairs through
    module procedure dot_pairs ! a specially named operator
  end interface 
contains
  integer function dot_pairs(pair1, pair2)
    type(pair), intent(in) :: pair1, pair2
    dot_pairs = pair1%first*pair2%first + pair1%second*pair2%second
  end function dot_pairs
end module
\end{lstlisting}
given which we can then write
\begin{lstlisting}
  integer    :: i
  type(pair) :: pair1, pair2
  [... some code to set up pair values ...]
  ! now multiply them
  i = pair1 * pair2
  i = pair1 .dot. pair2  ! equivalent to previous statement
\end{lstlisting}
Since the the multiplication operator (\texttt{*}) already exists for
all the default types, by defining it for a new type we have
\emph{overloaded} it. Note that there are some subtleties with
precedences of user-defined operators: operators (like \texttt{*})
that already exist have the same precedence as they have is usual
operators; operators that do not exist by default (\texttt{.dot}) have
the lowest possible preference, so, given the above definitions,
\begin{lstlisting}
  i = 2 + pair1 * pair2       ! legal
  i = 2 + pair1 .dot. pair2   ! illegal, means: (2+pair1).dot.pair2
  i = 2 + (pair1 .dot. pair2) ! legal
\end{lstlisting}
where the second line is illegal because we have not defined any
operator for adding an integer and a pair. Similarly care is needed
when using the \hoppet's operator \texttt{.conv.}.

\paragraph{Floating point precision:}
A final point concerns floating point variable types. Throughout we
have used definitions such as
\begin{lstlisting}
  real(dp), pointer :: pdf(:,:)
\end{lstlisting}
and written numbers with a trailing \texttt{\_dp}
\begin{lstlisting}
  param = 1.7_dp
\end{lstlisting}
Here \texttt{dp} is an integer parameter (defined in the
\texttt{types} module and accessible also through the
\texttt{hoppet\_v1} module), which specifies the \texttt{kind} of real
that we want to define, specifically double precision. We could also
have written \texttt{double precision} everywhere, but this is less
compact, and the use of a kind parameter has the advantage that we
can just modify its definition in one point in the program and the
precision will be modified everywhere. (Well, almost, since some
special functions are written in Fortran~77 using \texttt{double
  precision} declarations and do their numerics based on the
assumption that that truly is the type they're dealing with).

\paragraph{Optional and keyword arguments}

A feature of F95 that helps simplify user interfaces is that of
optional and keyword arguments. Suppose we have
\begin{lstlisting}
  subroutine hello(name, prefix, count)
    character(len=*),  intent(in)  :: name, prefix
    integer, optional, intent(in)  :: count
  end subroutine hello
\end{lstlisting}
Here the \ttt{count} argument is \ttt{optional} meaning that it need
not be supplied --- if it is absent the subroutine is should behave
sensibly all the same. Thus one can call the subroutine as
\begin{lstlisting}
  call hello(name, prefix)
  call hello(name, prefix, count)
\end{lstlisting}
Keyword arguments are useful if one doesn't want to remember the exact
order of a long list of arguments (or if one wants to specify just one
of several optional arguments). For example
\begin{lstlisting}
  call hello(name=name, prefix=prefix)
  call hello(prefix=prefix, name=name)
\end{lstlisting}
will do the same thing.

%=============================================================

%=====================================================================


\begin{thebibliography}{99}

\bibitem{Botje}
  M.~Botje, QCDNUM, \url{http://www.nikhef.nl/~h24/qcdnum/}~.

% Uses decomposition on Laguerre polynomials -- about
% 30 of them, remains Y^2 * T method. Initialisation
% (transform of splitting functions takes 15s on thalie)
% (didn't try evolution; didn't check accuracy; evolution
% times and accuracy are not mentioned; seemed fixed nf)
\bibitem{Schoeffel:1998tz}
L.~Schoeffel,
%``An elegant and fast method to solve QCD evolution equations,  application to
%the determination of the gluon content of the pomeron,''
Nucl.\ Instrum.\ Meth.\ A {\bf 423} (1999) 439.
%%CITATION = NUIMA,A423,439;%%
See also \url{http://www.desy.de/~schoffel/L_qcd98.html},
\url{http://www-spht.cea.fr/pisp/gelis/Soft/DGLAP/index.html}

\bibitem{Pegasus}
  A.~Vogt,
  %``Efficient evolution of unpolarized and polarized parton distributions  with
  %QCD-PEGASUS,''
  Comput.\ Phys.\ Commun.\  {\bf 170} (2005) 65
  [arXiv:hep-ph/0408244].
  %%CITATION = HEP-PH 0408244;%%

\bibitem{Pascaud:2001bi}
C.~Pascaud and F.~Zomer,
%``A fast and precise method to solve the Altarelli-Parisi equations in x
%space,''
arXiv:hep-ph/0104013.
%%CITATION = HEP-PH 0104013;%%


\bibitem{Weinzierl:2002mv}
S.~Weinzierl,
%``Fast evolution of parton distributions,''
Comput.\ Phys.\ Commun.\  {\bf 148} (2002) 314
[arXiv:hep-ph/0203112];
%%CITATION = HEP-PH 0203112;%%
%\bibitem{Roth:2004ti}
M.~Roth and S.~Weinzierl,
%``QED corrections to the evolution of parton distributions,''
Phys.\ Lett.\ B {\bf 590} (2004) 190
[arXiv:hep-ph/0403200].
%%CITATION = HEP-PH 0403200;%%

% about 1 minute at NLO.
\bibitem{coriano} A.~Cafarella and C.~Coriano,
%``Direct solution of renormalization group equations of QCD in x-space: NLO
%implementations at leading twist,''
Comput.\ Phys.\ Commun.\  {\bf 160} (2004) 213
[arXiv:hep-ph/0311313];
%%CITATION = HEP-PH 0311313;%%
%\bibitem{Cafarella:2005zj}
  A.~Cafarella, C.~Coriano' and M.~Guzzi,
  %``NNLO logarithmic expansions and exact solutions of the DGLAP equations from
  %x-space: New algorithms for precision studies at the LHC,''
  Nucl.\ Phys.\  B {\bf 748} (2006) 253
  [arXiv:hep-ph/0512358];
  %%CITATION = NUPHA,B748,253;%%
  A.~Cafarella, C.~Coriano and M.~Guzzi,
  %``Precision Studies of the NNLO DGLAP Evolution at the LHC with CANDIA,''
  arXiv:0803.0462 [hep-ph].

\bibitem{GuzziThesis}
  M.~Guzzi, Ph.D. Thesis, Lecce University, 2006 [hep-ph/0612355].

\bibitem{nnpdf}
  L.~Del Debbio, S.~Forte, J.~I.~Latorre, A.~Piccione and J.~Rojo  [NNPDF
                  Collaboration],
  %``Neural network determination of parton distributions: The nonsinglet
  %case,''
  JHEP {\bf 0703} (2007) 039
  [arXiv:hep-ph/0701127].

\bibitem{Kosower:1997hg}
  D.~A.~Kosower,
  %``Evolution of parton distributions,''
  Nucl.\ Phys.\  B {\bf 506} (1997) 439
  [arXiv:hep-ph/9706213].

\bibitem{Ratcliffe:2000kp}
  P.~G.~Ratcliffe,
  %``A matrix approach to numerical solution of the DGLAP evolution
  %equations,''
  Phys.\ Rev.\  D {\bf 63}, 116004 (2001)
  [arXiv:hep-ph/0012376].
  %%CITATION = PHRVA,D63,116004;%%

\bibitem{DGLAP}
V.N.~Gribov and L.N.~Lipatov, 
%\sjnp{15}{1972}{438};
Sov.\ J.\ Nucl.\ Phys. {\bf 15} (1972) 438;
%``Deep Inelastic E P Scattering In Perturbation Theory,''
%[Sov.\ J.\ Nucl.\ Phys.\  {\bf 15} (1972) 438].
%%CITATION = YAFIA,15,781;%%
G.~Altarelli and G.~Parisi, 
%\npb{126}{1977}{298};
Nucl.\ Phys.\ B {\bf 126} (1977) 298;
%``Asymptotic Freedom In Parton Language,''
%%CITATION = NUPHA,B126,298;%%
Yu.L.~Dokshitzer, 
%\jetp{46}{1977}{641}.
Sov.\ Phys.\ JETP {\bf 46} (1977) 641.
%``Calculation Of The Structure Functions For Deep Inelastic Scattering And E+ E- Annihilation By Perturbation Theory In Quantum Chromodynamics.
%[Zh.\ Eksp.\ Teor.\ Fiz.\  {\bf 73} (1977) 1216].
%%CITATION = SPHJA,46,641;%%

\bibitem{CTEQ}
  J.~Pumplin, D.~R.~Stump, J.~Huston, H.~L.~Lai, P.~Nadolsky and W.~K.~Tung,
  %``New generation of parton distributions with uncertainties from global  QCD
  %analysis,''
  JHEP {\bf 0207}, 012 (2002)
  [arXiv:hep-ph/0201195].

%\cite{Martin:2002dr}
\bibitem{MRST}
  A.~D.~Martin, R.~G.~Roberts, W.~J.~Stirling and R.~S.~Thorne,
  %``NNLO global parton analysis,''
  Phys.\ Lett.\  B {\bf 531} (2002) 216
  [arXiv:hep-ph/0201127].
  %%CITATION = PHLTA,B531,216;%%

\bibitem{DisResum}
  M.~Dasgupta and G.~P.~Salam,
  %``Resummation of the jet broadening in DIS,''
  Eur.\ Phys.\ J.\  C {\bf 24} (2002) 213
  [arXiv:hep-ph/0110213];
  %%CITATION = EPHJA,C24,213;%%
%\bibitem{Dasgupta:2002dc}
  %M.~Dasgupta and G.~P.~Salam,
  %``Resummed event-shape variables in DIS,''
  JHEP {\bf 0208} (2002) 032
  [arXiv:hep-ph/0208073].
  %%CITATION = JHEPA,0208,032;%%



\bibitem{caesar}
  A.~Banfi, G.~P.~Salam and G.~Zanderighi,
  %``Principles of general final-state resummation and automated
  %implementation,''
  JHEP {\bf 0503}, 073 (2005)
  [arXiv:hep-ph/0407286];
  %%CITATION = JHEPA,0503,073;%%
%
%\cite{Banfi:2004nk}
%\bibitem{Banfi:2004nk}
%  A.~Banfi, G.~P.~Salam and G.~Zanderighi,
  %``Resummed event shapes at hadron hadron colliders,''
  JHEP {\bf 0408}, 062 (2004)
  [arXiv:hep-ph/0407287].
  %%CITATION = JHEPA,0408,062;%%

\bibitem{Ciafaloni:2003rd}
  M.~Ciafaloni, D.~Colferai, G.~P.~Salam and A.~M.~Stasto,
  %``Renormalisation group improved small-x Green's function,''
  Phys.\ Rev.\  D {\bf 68}, 114003 (2003)
  [arXiv:hep-ph/0307188].


\bibitem{APPL}
  T.~Carli, G.~P.~Salam and F.~Siegert,
  %``A posteriori inclusion of PDFs in NLO QCD final-state calculations,''
  :hep-ph/0510324;
  %%CITATION = HEP-PH/0510324;%%
  T.~Carli, D.~Clements, {\it et al.}, in preparation.

\bibitem{Banfi:2007gu}
  A.~Banfi, G.~P.~Salam and G.~Zanderighi,
  %``Accurate QCD predictions for heavy-quark jets at the Tevatron and LHC,''
  JHEP {\bf 0707} (2007) 026
  [arXiv:0704.2999 [hep-ph]].


\bibitem{Benchmarks} 
  W.~Giele {\it et al.},
  ``Les Houches 2001, the QCD/SM working group: Summary report,''
  hep-ph/0204316, section 1.3;\\
  %%CITATION = HEP-PH 0204316;%%
  M.~Dittmar {\it et al.},
  ``Parton distributions: Summary report for the HERA-LHC workshop,''
  hep-ph/0511119, section 4.4.
  %%CITATION = HEP-PH 0511119;%%

\bibitem{LHAPDF} W.~Giele and M.~R.~Whalley,
\url{http://hepforge.cedar.ac.uk/lhapdf/}


%\cite{Moch:2004pa}
\bibitem{NNLO-NS}
  S.~Moch, J.~A.~M.~Vermaseren and A.~Vogt,
  %``The three-loop splitting functions in QCD: The non-singlet case,''
  Nucl.\ Phys.\ B {\bf 688} (2004) 101
  [arXiv:hep-ph/0403192].
  %%CITATION = HEP-PH 0403192;%%

%\cite{Vogt:2004mw}
\bibitem{NNLO-singlet}
  A.~Vogt, S.~Moch and J.~A.~M.~Vermaseren,
  %``The three-loop splitting functions in QCD: The singlet case,''
  Nucl.\ Phys.\ B {\bf 691} (2004) 129
  [arXiv:hep-ph/0404111].
  %%CITATION = HEP-PH 0404111;%%

%\cite{Dokshitzer:2005bf}
\bibitem{Dokshitzer:2005bf}
  Yu.~L.~Dokshitzer, G.~Marchesini and G.~P.~Salam,
  %``Revisiting parton evolution and the large-x limit,''
  Phys.\ Lett.\  B {\bf 634}, 504 (2006)
  [arXiv:hep-ph/0511302].
  %%CITATION = PHLTA,B634,504;%%


\bibitem{Mitov:2006ic}
  A.~Mitov, S.~Moch and A.~Vogt,
  %``Next-to-next-to-leading order evolution of non-singlet fragmentation
  %functions,''
  Phys.\ Lett.\  B {\bf 638} (2006) 61
  [arXiv:hep-ph/0604053].

\bibitem{Basso:2006nk}
  B.~Basso and G.~P.~Korchemsky,
  %``Anomalous dimensions of high-spin operators beyond the leading order,''
  Nucl.\ Phys.\  B {\bf 775} (2007) 1
  [arXiv:hep-th/0612247].
  %%CITATION = NUPHA,B775,1;%%

\bibitem{Dokshitzer:2006nm}
  Yu.~L.~Dokshitzer and G.~Marchesini,
  %``N = 4 SUSY Yang-Mills: Three loops made simple(r),''
  Phys.\ Lett.\  B {\bf 646} (2007) 189
  [arXiv:hep-th/0612248].

\bibitem{Beccaria:2007bb}
  M.~Beccaria, Yu.~L.~Dokshitzer and G.~Marchesini,
  %``Twist 3 of the sl(2) sector of N=4 SYM and reciprocity respecting
  %evolution,''
  Phys.\ Lett.\  B {\bf 652} (2007) 194
  [arXiv:0705.2639 [hep-th]].


\bibitem{NNLO-MTM}
  M.~Buza, Y.~Matiounine, J.~Smith, R.~Migneron and W.~L.~van Neerven,
  %``Heavy quark coefficient functions at asymptotic values $Q~2 \gg m~2$,''
  Nucl.\ Phys.\ B {\bf 472}, 611 (1996)
  [arXiv:hep-ph/9601302];\\
  %%CITATION = HEP-PH 9601302;%%
%
  M.~Buza, Y.~Matiounine, J.~Smith and W.~L.~van Neerven,
  %``Charm electroproduction viewed in the variable-flavour number scheme
  %versus fixed-order perturbation theory,''
  Eur.\ Phys.\ J.\ C {\bf 1}, 301 (1998)
  [arXiv:hep-ph/9612398].
  %%CITATION = HEP-PH 9612398;%%

\bibitem{Chetyrkin:1997sg}
  K.~G.~Chetyrkin, B.~A.~Kniehl and M.~Steinhauser,
  %``Strong coupling constant with flavour thresholds at four loops in the
  %MS-bar scheme,''
  Phys.\ Rev.\ Lett.\  {\bf 79}, 2184 (1997)
  [arXiv:hep-ph/9706430].

\bibitem{vanNeerven:1999ca}
  W.~L.~van Neerven and A.~Vogt,
  %``NNLO evolution of deep-inelastic structure functions: The non-singlet
  %case,''
  Nucl.\ Phys.\ B {\bf 568} (2000) 263
  [arXiv:hep-ph/9907472].
  %%CITATION = HEP-PH 9907472;%%

\bibitem{vanNeerven:2000uj}
  W.~L.~van Neerven and A.~Vogt,
  %``NNLO evolution of deep-inelastic structure functions: The singlet case,''
  Nucl.\ Phys.\ B {\bf 588} (2000) 345
  [arXiv:hep-ph/0006154].
  %%CITATION = HEP-PH 0006154;%%

\bibitem{NRf90}
  Press {\it et al.}, \emph{Numerical Recipes in Fortran~90},
  Cambridge University Press, 1996.
  
\bibitem{VogtMTMParam} A.~Vogt, private communication.


\bibitem{White:2005wm}
  C.~D.~White and R.~S.~Thorne,
  %``Comparison of NNLO DIS scheme splitting functions with results from exact
  %gluon kinematics at small x,''
  Eur.\ Phys.\ J.\ C {\bf 45} (2006) 179
  [arXiv:hep-ph/0507244].
  %%CITATION = HEP-PH 0507244;%%

%\cite{Bethke:2006ac}
\bibitem{Bethke:2006ac}
  S.~Bethke,
  %``Experimental tests of asymptotic freedom,''
  Prog.\ Part.\ Nucl.\ Phys.\  {\bf 58}, 351 (2007)
  [arXiv:hep-ex/0606035].
  %%CITATION = PPNPD,58,351;%%

%\cite{de Florian:2007hc}
\bibitem{de Florian:2007hc}
  D.~de Florian, R.~Sassot and M.~Stratmann,
  %``Global Analysis of Fragmentation Functions for Protons and Charged
  %Hadrons,''
  Phys.\ Rev.\  D {\bf 76}, 074033 (2007)
  [arXiv:0707.1506 [hep-ph]].
  %%CITATION = PHRVA,D76,074033;%%


\bibitem{Corcella:2005us}
  G.~Corcella and L.~Magnea,
  %``Soft-gluon resummation effects on parton distributions,''
  Phys.\ Rev.\  D {\bf 72} (2005) 074017
  [arXiv:hep-ph/0506278].

\bibitem{Martin:2007bv}
  A.~D.~Martin, W.~J.~Stirling, R.~S.~Thorne and G.~Watt,
  %``Update of Parton Distributions at NNLO,''
  Phys.\ Lett.\  B {\bf 652}, 292 (2007)
  [arXiv:0706.0459 [hep-ph]].

%\cite{Tung:2006tb}
\bibitem{Tung:2006tb}
  W.~K.~Tung, H.~L.~Lai, A.~Belyaev, J.~Pumplin, D.~Stump and C.~P.~Yuan,
  %``Heavy quark mass effects in deep inelastic scattering and global QCD
  %analysis,''
  JHEP {\bf 0702}, 053 (2007)
  [arXiv:hep-ph/0611254].
  %%CITATION = JHEPA,0702,053;%%

\bibitem{Martin:2004dh}
  A.~D.~Martin, R.~G.~Roberts, W.~J.~Stirling and R.~S.~Thorne,
  %``Parton distributions incorporating QED contributions,''
  Eur.\ Phys.\ J.\  C {\bf 39}, 155 (2005)
  [arXiv:hep-ph/0411040].

\bibitem{Ciafaloni:2000df}
  M.~Ciafaloni, P.~Ciafaloni and D.~Comelli,
  %``Bloch-Nordsieck violating electroweak corrections to inclusive TeV  scale
  %hard processes,''
  Phys.\ Rev.\ Lett.\  {\bf 84}, 4810 (2000)
  [arXiv:hep-ph/0001142].


\bibitem{Ciafaloni:2005fm}
  P.~Ciafaloni and D.~Comelli,
  %``Electroweak evolution equations,''
  JHEP {\bf 0511}, 022 (2005)
  [arXiv:hep-ph/0505047].
  %%CITATION = JHEPA,0511,022;%%


\bibitem{FortranPolyLog}
  T.~Gehrmann and E.~Remiddi,
  %``Numerical evaluation of two-dimensional harmonic polylogarithms,''
  Comput.\ Phys.\ Commun.\  {\bf 144} (2002) 200.
  %[hep-ph/0111255].
  %%CITATION = HEP-PH 0111255;%%

\bibitem{F95Explained}
  M. Metcalf and J. Reid, \emph{Fortran 90/95 Explained}, Oxford
  University Press, 1996.

\bibitem{F95WebResources} Many introductions and tutorials about
  fortran~90 may be found at
  \url{http://dmoz.org/Computers/Programming/Languages/Fortran/Tutorials/Fortran_90_and_95/}

\end{thebibliography}
\end{document}